\documentclass[10pt]{article}
\usepackage{multicol}
\usepackage{graphicx}
\usepackage{amsmath}

\begin{document}

\begin{center}
{\large \bf Examining the model dependence of the determination of
kinetic freeze-out temperature and transverse flow velocity in
small collision system}

\vskip.75cm

Hai-Ling Lao$^1$, Fu-Hu Liu$^{1,}${\footnote{E-mail:
fuhuliu@163.com; fuhuliu@sxu.edu.cn}}, Bao-Chun Li$^1$, Mai-Ying
Duan$^1$, Roy A. Lacey$^2$

{\small\it $^1$Institute of Theoretical Physics \& State Key
Laboratory of Quantum Optics and Quantum Optics Devices,\\ Shanxi
University, Taiyuan, Shanxi 030006, China

$^2$Departments of Chemistry \& Physics, Stony Brook University,
Stony Brook, NY 11794, USA}
\end{center}

\vskip.75cm

{\bf Abstract:} The transverse momentum distributions of the
identified particles produced in small collision systems at the
Relativistic Heavy Ion Collider (RHIC) and Large Hadron Collider
(LHC) have been analyzed by four models. The first two models
utilize the blast-wave model with different statistics. The last
two models employ certain linear correspondences based on
different distributions. The four models describe the experimental
data measured by the Pioneering High Energy Nuclear Interaction
eXperiment (PHENIX), Solenoidal Tracker at RHIC (STAR), and A
Large Ion Collider Experiment (ALICE) cCollaborations equally
well. It is found that both the kinetic freeze-out temperature and
transverse flow velocity in the central collisions are comparable
with those in the peripheral collisions. With the increase of
collision energy from that of the RHIC to that of the LHC, the
considered quantities typically do not decrease. Comparing with
the central collisions, the proton-proton collisions are closer to
the peripheral collisions.
\\

{\bf Keywords:} kinetic freeze-out temperature, transverse flow
velocity, small collision system, central collisions, peripheral
collisions
\\

{\bf PACS:} 25.75.Ag, 25.75.Dw, 24.10.Pa

\vskip1.0cm

\begin{multicols}{2}

{\section{Introduction}}

As an important concept in both thermal and subatomic physics,
temperature is widely used in experimental measurements and
theoretical studies. Contrary to macroscopic thermal physics,
temperature in microscopic subatomic physics cannot be measured
directly; nevertheless, the temperature measured in thermal
physics is manifested by the change of a given quantity of the
thermometric material. Instead, we can calculate the temperature
by using the methods of particle ratios and transverse momentum
($p_T$) spectra. The temperature obtained from particle ratios is
typically the chemical freeze-out temperature ($T_{ch}$), which
can describe the degree of excitation of the interacting system at
the stage of chemical equilibrium. The temperature obtained from
the $p_T$ spectra with a thermal distribution that does not
include the flow effect, is typically an effective temperature
($T_{eff}$ or $T$) which is not a real temperature due to its
relation to particle mass. The temperature obtained from $p_T$
spectra with the thermal distribution which includes flow effect
is usually the kinetic freeze-out temperature ($T_{kin}$ or $T_0$)
which describes the degree of excitation of the interacting system
at the stage of kinetic and thermal equilibrium.

The chemical freeze-out and kinetic freeze-out are two main stages
of the evolution of the interacting system in high energy
collisions. At the stage of chemical freeze-out, the chemical
components (relative fractions) of the particles are fixed. At the
stage of kinetic freeze-out, the $p_T$ and momentum ($p$) spectra
of the particles are no longer changed. We are interested in the
$T_0$ value, owing to its relation to the $p_T$ spectrum of the
identified particles, which is one of the quantities measured
first in our experiments. At the same time, $T_0$ is related to
the structure of the phase diagram in the $T_0$-related spaces,
such as $T_0$ as a function of $\beta_T$ and as a function of
$\sqrt{s_{NN}}$, where $\beta_T$ is the mean transverse flow
velocity, resulted from the impact and squeeze while
$\sqrt{s_{NN}}$ denotes the center-of-mass energy per nucleon pair
in collisions of nuclei [$\sqrt{s}$ in particle collisions such as
in proton-proton ($p$-$p$ or $pp$) collisions]. In particular, in
the energy ranges available in the beam energy scan (BES) program
at the Relativistic Heavy Ion Collider (RHIC) and the BES program
at the Super Proton Synchrotron (SPS), the chemical potential
($\mu_B$) of baryons needs to be considered. Then, the structure
of phase diagram in the $T_0$ versus $\mu_B$ space can be studied
in both the RHIC BES and the SPS BES energy ranges.

Generally, $\mu_B$ can be obtained from the particle ratios and
its excitation function has been studied in detail [1--5], while
$T_0$ and $\beta_T$ can be obtained from the $p_T$ spectra. In
Refs. [6--13], different methods have been used to obtain $T_0$
and $\beta_T$. In our recent studies [14--17], we have used a
number of models to obtain $T_0$ and $\beta_T$ in nucleus-nucleus
[gold-gold (Au-Au) and lead-lead (Pb-Pb)] collisions at the RHIC
and Large Hadron Collider (LHC) energies, where the top RHIC
energy was $\sqrt{s_{NN}}=200$ GeV while the LHC energy reached a
few TeV. Similar results were obtained when a non-zero $\beta_T$
was used in peripheral nucleus-nucleus collisions in the
Blast-Wave model with Boltzmann-Gibbs statistics (BGBW model)
[6--8, 18] and with Tsallis statistics (TBW model) [9, 18, 19].
Our results show that $T_0$ ($\beta_T$) in central nucleus-nucleus
collisions is comparable to that in peripheral collisions.
Similarly, the values of $T_0$ and $\beta_T$ at the LHC are close
to those at the RHIC.

It is interesting to compare the results of different models in
small collision systems such as $pp$ and deuteron-gold ($d$-Au)
collisions at the RHIC, and $pp$ and proton-lead ($p$-Pb)
collisions at the LHC. In this paper, we use four models to obtain
$T_0$ and $\beta_T$ values from the $p_T$ spectra of the
identified particles produced in $pp$ and $d$-Au collisions at the
RHIC, and in $pp$ and $p$-Pb collisions at the LHC. The model
results of the $p_T$ spectra are compared with each other and with
the experimental data of the Pioneering High Energy Nuclear
Interaction eXperiment (PHENIX) [20], Solenoidal Tracker at RHIC
(STAR) [21--23], and A Large Ion Collider Experiment (ALICE)
collaborations [24--25]. Then, similar $T_0$ and $\beta_T$ values
are obtained from the analyses of the experimental data by the
four models.

The paper is structured as follows. The formalism and method are
described in Section 2. Results and discussion are given in
Section 3. In Section 4, we summarize our main observations and
conclusions.
\\

{\section{Formalism and method}}

In the present work, four models were used for the $p_T$
distributions for comparisons in small collision systems;
nevertheless, in our recent work [14] they were employed to obtain
$T_0$ and $\beta_T$ values in nucleus-nucleus collisions at RHIC
and LHC energies using a different superposition of soft
excitation and hard scattering components. In order to provide a
comprehensive review of the present work, we discuss the previous
studies of the four models as follows.

i) BGBW model [6--8]: in this model we considered a non-zero
$\beta_T$ of the produced particles.

According to refs. [6--8], the BGBW model gives the $p_T$
distribution as
\begin{align}
f_1(p_T)&=\frac{1}{N}\frac{dN}{dp_T} =C_1 p_T m_T \int_0^R rdr \times \nonumber\\
& I_0 \bigg[\frac{p_T \sinh(\rho)}{T_0} \bigg] K_1 \bigg[\frac{m_T
\cosh(\rho)}{T_0} \bigg],
\end{align}
where $N$ is the number of particles, $C_1$ is a normalized
constant, $I_0$ and $K_1$ are modified Bessel functions of the
first and second kinds, respectively, $m_T=\sqrt{p_T^2+m_0^2}$ is
the transverse mass, $\rho= \tanh^{-1} [\beta(r)]$ is the boost
angle, $\beta(r)= \beta_S(r/R)^{n_0}$ is a self-similar flow
profile, $\beta_S$ is the flow velocity on the surface, $r/R$ is
the relative radial position in the thermal source [6], and
$n_0=2$ similarly to that in ref. [6]. The relation between
$\beta_T$ and $\beta(r)$ is $\beta_T= (2/R^2)\int_0^R r\beta(r)dr
=2\beta_S/(n_0+2)=0.5\beta_S$.

ii) TBW model [9]: in this model we also considered a non-zero
$\beta_T$.

According to refs. [9], the TBW model gives the $p_T$ distribution
in the form of
\begin{align}
f_2(p_T)&=\frac{1}{N}\frac{dN}{dp_T} =C_2 p_T m_T \int_{-\pi}^{\pi} d\phi \int_0^R rdr \Big\{1+ \nonumber\\
& \frac{q-1}{T_0} \big[ m_T \cosh(\rho) -p_T \sinh(\rho)
\cos(\phi)\big] \Big\}^{-q/(q-1)},
\end{align}
where $C_2$ is a normalized constant, $q$ is an entropy index
characterizing the degree of non-equilibrium, $\phi$ denotes the
azimuth [9], and $n_0=1$ similarly to that in ref. [9]. In the
first two models, $n_0$ is independent: it does not matter if
$n_0=1$ or $n_0=2$ is used. To be compatible with refs. [6] and
[9], we use $n_0=2$ in the first model and $n_0=1$ in the second
model. It should be noted that we use the index $-q/(q-1)$ in Eq.
(2) instead of $-1/(q-1)$ in ref. [9] due to the fact that $q$ is
very close to one. This substitution results in a small and
negligible difference in the Tsallis distribution [19].

iii) An alternative method, in which the intercept in the $T$
versus $m_0$ relation is assumed to be $T_0$ [7, 10--13], the
slope in the $\langle p_T \rangle$ versus $\overline{m}$ relation
is assumed to be $\beta_T$, and the slope in the $\langle p
\rangle$ versus $\overline{m}$ relation is assumed to be the
radial flow velocity $\beta$ [14--17], which does not include the
contribution of longitudinal flow. Here $m_0$ denotes the rest
mass, $\overline{m}$ denotes the mean moving mass (mean energy),
$\langle ... \rangle$ denotes the theoretical distribution average
of the considered quantity, and $T$ is obtained from a Boltzmann
distribution [18].

Two steps are required to obtain $T_0$ and $\beta_T$. To use the
relations $T=T_0+am_0$, $\langle p_T \rangle =b_1+\beta_T
\overline{m}$, and $\langle p \rangle=b_2+ \beta \overline{m}$,
where $a$, $b_1$, and $b_2$ are fitted parameters, we choose the
form of Boltzmann distribution as [18]
\begin{align}
f_{3}(p_T)=\frac{1}{N}\frac{dN}{dp_T} =C_{3} p_T m_T \exp
\bigg(-\frac{m_T}{T} \bigg),
\end{align}
where $C_{3}$ is a normalized constant related to the free
parameter $T$ and particle mass $m_0$ via its relation to $m_T$;
nevertheless, the Boltzmann distribution has multiple forms [18].

iv) This model is similar to the third model, but $T$ is obtained
from a Tsallis distribution [18, 19].

We choose the Tsallis distribution in the form of [18, 19]
\begin{align}
f_{4}(p_T)=\frac{1}{N}\frac{dN}{dp_T} =C_{4} p_T m_T \bigg(1+
\frac{q-1}{T} m_T \bigg)^{-q/(q-1)},
\end{align}
where $C_{4}$ is a normalized constant related to the free
parameters $T$ and $q$, as well as $m_0$; nevertheless, the
Tsallis distribution has more than one forms [18, 19].

Similarly to our recent work [14], in both the BGBW and TBW
models, a non-zero $\beta_T$ of the produced particles is
considered in the peripheral nucleus-nucleus collisions. The
peripheral collisions contain a small number of participant
nucleons that take part in the violent interactions. This
condition is similar to a small collision system, which also
contains a small number of participant nucleons. When the cold
nuclear effect is neglected, the small collision system is similar
to a peripheral collisions. This means that a non-zero $\beta_T$
needs to be considered for the small collision system to maintain
consistency; however, the values of $\beta_T$ for a small
collision system and peripheral collisions are possibly different.
Naturally, it is not unusual if the values of $\beta_T$ in the two
types of collisions are nearly the same.

From the first model $T_0$ and $\beta_T$ can be obtained, while
from the second model $T_0$, $\beta_T$, and $q$ can be obtained.
The first two models are employed to compare their results.
Although the forms of the first two models are obviously
different, the values of $T_0$ ($\beta_T$) obtained from them
exhibit a little difference only. The last two models are used for
comparison as well. The obtained values of the last two models
exhibit a little difference as well; however they are still
noticeably different.

The description of the above models is presented at mid-rapidity,
in which $y \approx 0$, where $y\equiv 0.5\ln[(E+p_z)/(E-p_z)]$,
and $E$ and $p_z$ denote the energy and longitudinal momentum,
respectively. At high $p_T$, $y\approx -\ln\tan(\vartheta/2)\equiv
\eta$, where $\vartheta$ and $\eta$ denote the emission angle and
pseudorapidity of the considered particle, respectively. The
effect of the spin and chemical potential on the $p_T$ spectra is
neglected because they are small at the top RHIC and LHC energies
[1--4]. Similarly to our recent work [14], the kinetic freeze-out
temperature, the mean transverse (radial) flow velocity, and the
effective temperature in different models are uniformly denoted by
$T_0$, $\beta_T$, and $T$, respectively; however, different values
can be obtained by different models.

Equations (1)--(4) are the functions describing mainly the
contribution of the soft excitation process. These are only valid
for the spectra in a narrow $p_T$ range, which mainly covers the
range mainly from 0 to 2.5--3.5 GeV/$c$ in most cases or a
slightly higher in certain cases. Even for the soft excitation
process, the Boltzmann distribution is not sufficient to fit the
$p_T$ spectra in certain cases. In the case of a two- or
three-component Boltzmann distribution, $T$ is the weighted
average resulting from different effective temperatures and the
corresponding fractions obtained from different components.

Generally, in the present work, two main processes in high energy
collisions are considered. Apart from the soft excitation process,
the main process is the hard scattering process, which contributes
to the spectra in a wide $p_T$ range and according to the quantum
chromodynamics (QCD) calculation [26--28], it can be described by
an inverse power-law as
\begin{align}
f_H(p_T)=\frac{1}{N}\frac{dN}{dp_T} =Ap_T \bigg( 1+\frac{p_T}{
p_0} \bigg)^{-n},
\end{align}
where $p_0$ and $n$ are free parameters, and $A$ is a normalized
constant related to the free parameters. As a result of the
QCD-based calculation, Eq. (5) contributes to the distribution in
a range of 0 to high $p_T$. Theoretically, in spite of the
overlapping regions in the low $p_T$ range between the
contributions of Eqs. (1)--(4) and (5), they cannot replace each
other.

The experimental $p_T$ spectra are typically distributed in a wide
range. This means that a superposition of both the contributions
of soft and hard processes (components) needs to be used to fit
the spectra. We use the usual step function for structuring the
superposition in order to avoid the entanglement between the
contribution ranges of the soft excitation and hard scattering
components, such that
\begin{align}
f_0(p_T)&=\frac{1}{N}\frac{dN}{dp_T}=A_1\theta(p_1-p_T)f_S(p_T) \nonumber\\
&+A_2\theta(p_T-p_1)f_H(p_T),
\end{align}
where $f_S(p_T)$ denotes one of Eqs. (1)--(4), $A_1$ and $A_2$ are
constants, ensuring that the contributions of soft and hard
components are the same at $p_T=p_1$, and the step function
$\theta(x)=1$ if $x>0$ and $\theta(x)=0$ if $x<0$. The fraction
(rate) of the contribution of the soft component is given by
$k=\int_0^{p_1}A_1f_S(p_T)dp_T$. Owing to the respective ranges of
the different contributions, the selection of parameters in Eqs.
(1)--(4) and (5) has no effect on their correlation and dependence
on each other.

In certain cases, the contribution of the resonance production for
pions and the strong stopping effect for the participating
nucleons are non-negligible at very low ranges. A very-soft
component needs to be used for the $p_T$ values ranging from 0 to
0.5--1.5 GeV/$c$. Let us consider the contribution of the
very-soft component. Equation (6) can be rewritten as
\begin{align}
f_0(p_T)&=\frac{1}{N}\frac{dN}{dp_T} =A_{VS}\theta(p_{VS}-p_T)f_{VS}(p_T) \nonumber\\
&+A_1\theta(p_T-p_{VS})\theta(p_1-p_T)f_S(p_T) \nonumber\\
&+A_2\theta(p_T-p_1)f_H(p_T),
\end{align}
where $f_{VS}(p_T)$ denotes one of Eqs. (1)--(4) similarly to
$f_S(p_T)$, and $A_{VS}$ is a constant ensuring that the
contributions of the very-soft and soft components are the same at
$p_T=p_{VS}$. Let us denote the rates of the very-soft and soft
components by $k_{VS}$ and $k_S$, respectively. Then,
$k_{VS}=\int_0^{p_{VS}}A_{VS} f_{VS}(p_T)dp_T$ and
$k_S=\int_{p_{VS}}^{p_1}A_1 f_S(p_T)dp_T$, where $k_{VS}+k_S=k$
[for the definition of $k$, please refer to the section following
Eq. (6)].

Although $f_{VS}(p_T)$ and $f_S(p_T)$ have the same form in Eq.
(7), their contribution ranges are different. Similarly, the
contribution range of $f_H(p_T)$ is different from those of
$f_{VS}(p_T)$ and $f_S(p_T)$. The three functions have no
correlation or dependence in the fitting procedure. We fitted
$f_{VS}(p_T)$ at very-soft $p_T$ ranging from 0 to 0.5--1.5
GeV/$c$, $f_S(p_T)$ at soft $p_T$ ranging from 0.5--1.5 GeV/$c$ to
2.5--3.5 GeV/$c$, and $f_H(p_T)$ at hard $p_T$ ranging from
2.5--3.5 GeV/$c$ to the maximum. In the case of without
$f_{VS}(p_T)$, Eq. (7) transforms into Eq. (6). Then, we fitted
$f_S(p_T)$ in Eq. (6) in the range of 0 to 2.5--3.5 GeV/$c$. In
the calculation, because of their different fractions, we used the
weighted average of parameters in very-soft and soft components in
Eq. (7) to compare them with the values obtained from Eqs. (6) and
(7).
\\

{\section{Results and discussion}}

\begin{figure*}[!htb]
\begin{center}
\includegraphics[width=16.0cm]{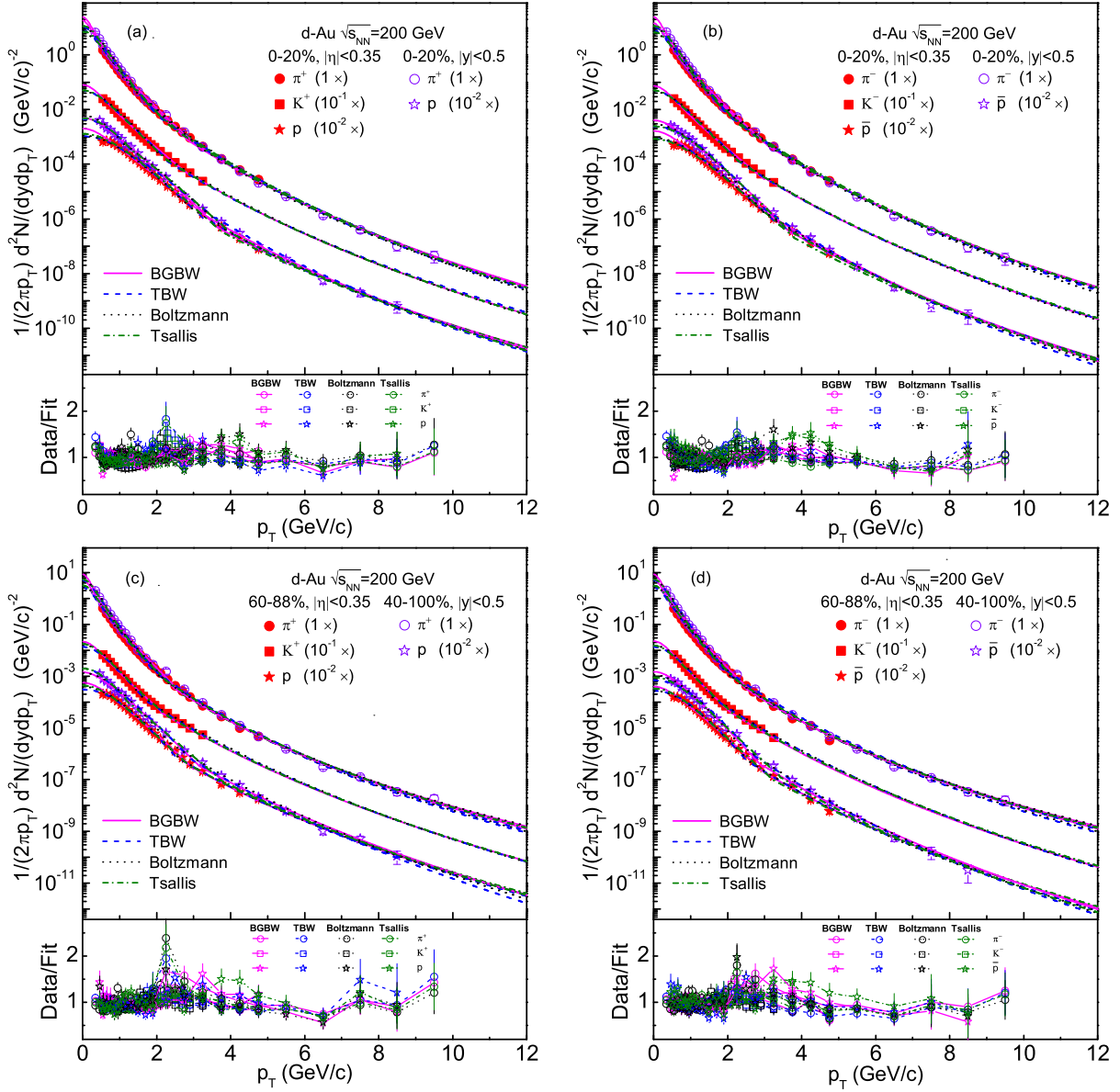}
\end{center} \vskip-.5cm
{\caption{\small (Color online) Transverse momentum spectra of
$\pi^+$, $K^+$, and $p$ [panels (a) and (c)], as well as $\pi^-$,
$K^-$, and $\bar p$ [panels (b) and (d)] produced in 0--20\%
[panels (a) and (b)] and 60--88\% (40--100\%) [panels (c) and (d)]
$d$-Au collisions at $\sqrt{s_{NN}}=200$ GeV. For clarity, the
spectra for different particles are multiplied by different
amounts shown in the panels. The closed and open symbols represent
the experimental data of the PHENIX and STAR collaborations
measured in $|\eta|<0.35$ [20] and $|y|<0.5$ [21], respectively.
The solid, dashed, dotted, and dashed-dotted curves are our
results fitted by Eqs. (6) and (7) in which $f_S(p_T)$
($f_{VS}(p_T)$) denote $f_1(p_T)$, $f_2(p_T)$, $f_3(p_T)$, and
$f_4(p_T)$, respectively. The bottom panels show the data for the
fitting of the ratios.}}
\end{figure*}

\begin{figure*}[!htb]
\begin{center}
\includegraphics[width=16.0cm]{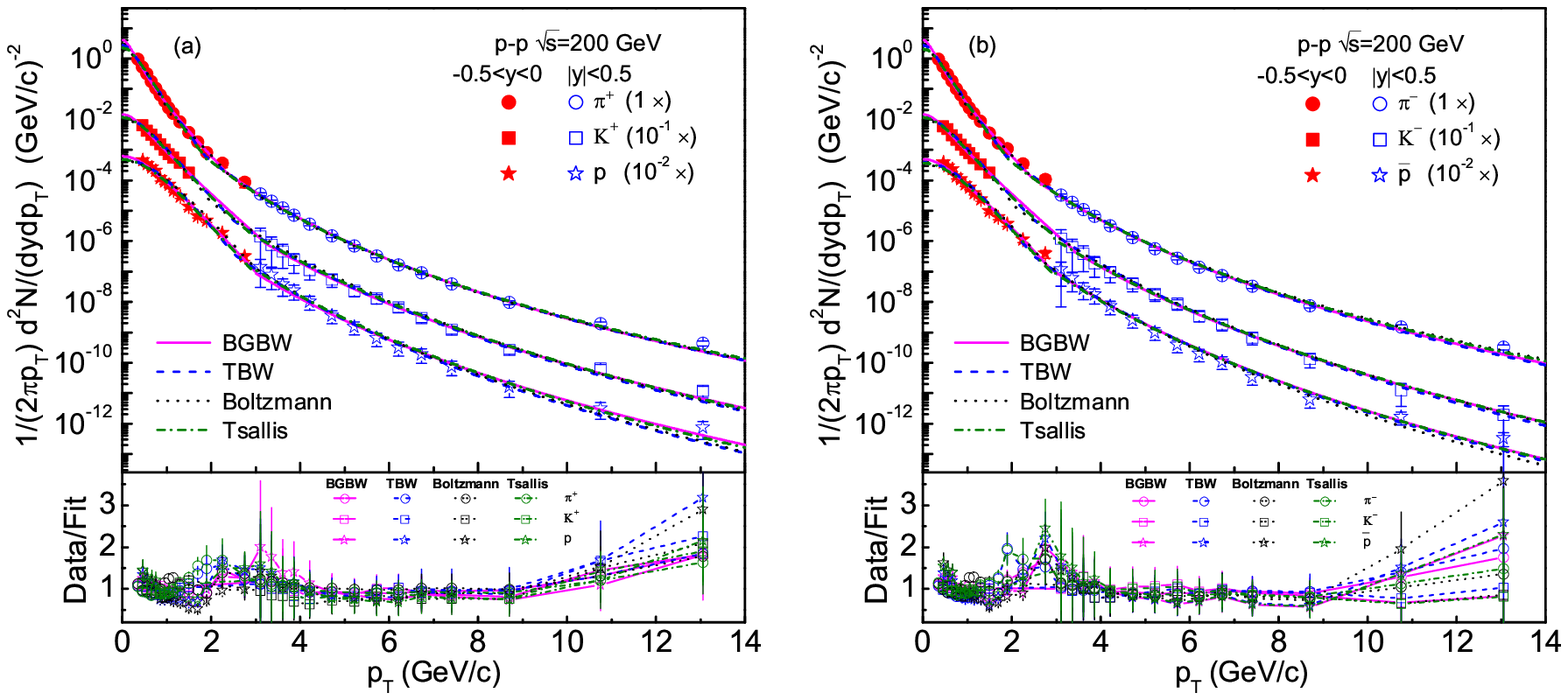}
\end{center} \vskip-.5cm
{\caption{\small (Color online) Spectra of $\pi^+$, $K^+$, and $p$
(panel (a)), as well as $\pi^-$, $K^-$, and $\bar p$ (panels (b)),
produced in $pp$ collisions at $\sqrt{s}=200$ GeV. The closed and
open symbols represent the experimental data of the STAR
collaboration measured in $-0.5<y<0$ and $|y|<0.5$, respectively
[22, 23].}}
\end{figure*}

In Fig. 1, the transverse momentum spectra, $1/(2\pi p_T) \cdot
d^2N/(dydp_T)$, are shown for positively charged pions ($\pi^+$),
positively charged kaons ($K^+$), and protons ($p$) [Figs. 1(a)
and 1(c)], as well as negatively charged pions ($\pi^-$),
negatively charged kaons ($K^-$), and antiprotons ($\bar p$)
[Figs. 1(b) and 1(d)] produced in 0--20\% [Figs. 1(a) and 1(b)]
and 60--88\% (40--100\%) [Figs. 1(c) and 1(d)] $d$-Au collisions
at $\sqrt{s_{NN}}=200$ GeV. The closed and open symbols represent
the experimental data of the PHENIX and STAR Collaboration
measured in the pseudorapidity range $|\eta|<0.35$ [20] and the
rapidity range $|y|<0.5$ [21], respectively. The curves show the
results obtained by models i)--iv) and the fit parameters are
given in Tables 1--4, respectively, with most of them are fitted
by Eq. (6). The numerical values fitted by Eq. (7) are marked by a
star at the end of the line, where the results obtained from the
very-soft and soft components are shown together. It can be seen
that the four considered models describe the $p_T$ spectra of the
identified particles produced in central (0--20\%) and peripheral
(60--88\% and 40--100\%) $d$-Au collisions at $\sqrt{s_{NN}}=200$
GeV similarly well.

Similarly to Fig. 1, Figs. 2(a) and 2(b) show the spectra of
$\pi^+$, $K^+$, and $p$, as well as $\pi^-$, $K^-$, and $\bar p$,
produced in $pp$ collisions at $\sqrt{s}=200$ GeV. The closed and
open symbols represent the experimental data of the STAR
collaboration measured in the range of $-0.5<y<0$ and at
$|y|<0.5$, respectively [22, 23]. The fitting parameters are given
in Tables 1--4. It can be seen that the four considered models
describe the $p_T$ spectra of the identified particles produced in
$pp$ collisions at $\sqrt{s}=200$ GeV similarly well.

Figure 3 is similar to Fig. 1, and it shows the spectra of
$\pi^++\pi^-$, $K^++K^-$, and $p+\bar p$ produced in 0--5\% [Fig.
3(a)] and 80--100\% [Fig. 3(b)] $p$-Pb collisions at
$\sqrt{s_{NN}}=5.02$ TeV. The symbols represent the experimental
data of the ALICE collaboration measured in the range of
$-0.5<y<0$ [24]. It can be seen in most cases that the four
considered models describe the $p_T$ spectra of the identified
particles produced in $p$-Pb collisions at $\sqrt{s_{NN}}=5.02$
TeV similarly well.

\begin{figure*}[!htb] \vskip.5cm
\begin{center}
\includegraphics[width=16.0cm]{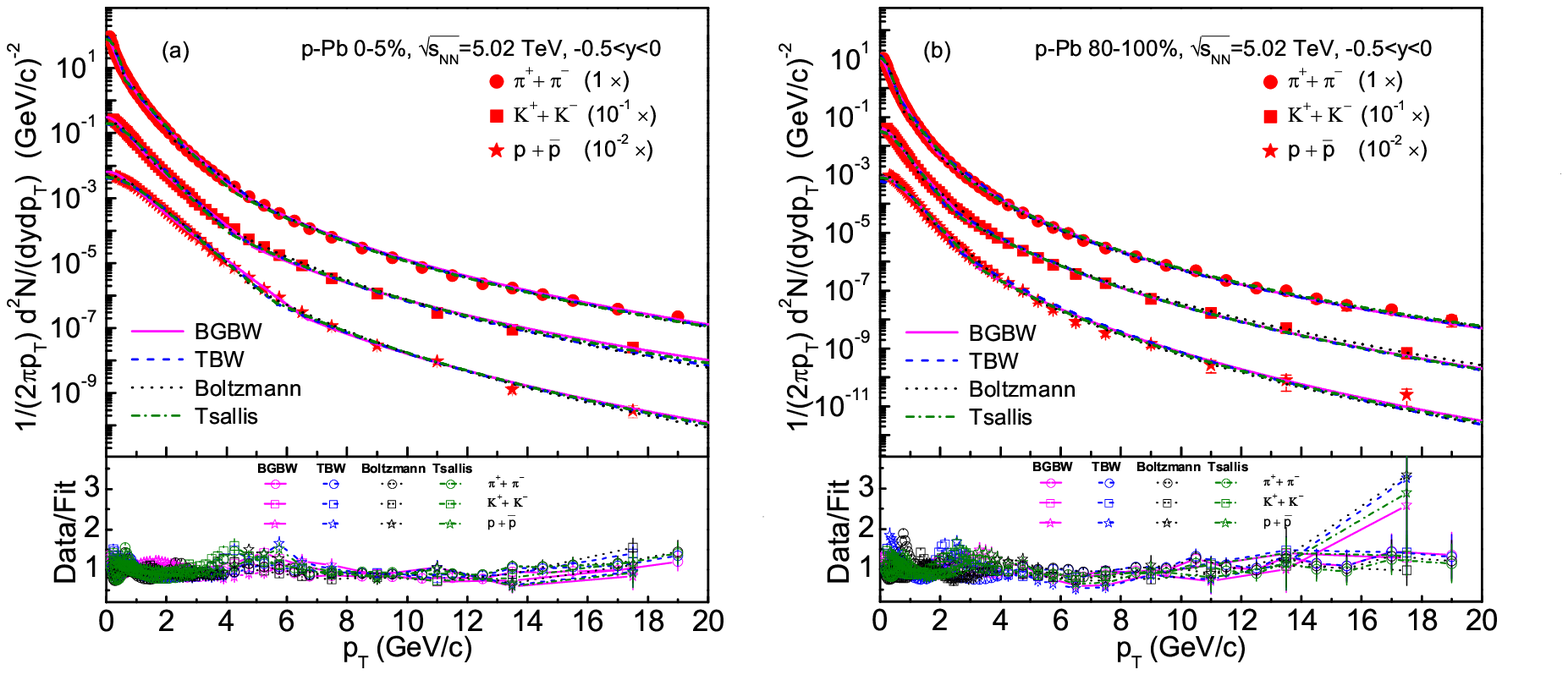}
\end{center} \vskip-.5cm
{\caption{\small (Color online) Spectra of $\pi^++\pi^-$,
$K^++K^-$, and $p+\bar p$ produced in 0--5\% [panel (a)] and
80--100\% [panel (b)] $p$-Pb collisions at $\sqrt{s_{NN}}=5.02$
TeV. The symbols represent the experimental data of the ALICE
collaboration measured in the range of $-0.5<y<0$ [24].}}
\end{figure*}

Similarly to Fig. 1, Fig. 4 shows spectra, $(1/N_{EV}) \cdot
1/(2\pi p_T) \cdot d^2N/(dydp_T)$, of $\pi^++\pi^-$, $K^++K^-$,
and $p+\bar p$ produced in $pp$ collisions at $\sqrt{s}=2.76$ TeV,
where $N_{EV}$ denotes the number of events and it is typically
omitted. The symbols represent the experimental data of the ALICE
collaboration measured in $|y|<0.5$ for low-$p_T$ particles and in
$|\eta|<0.8$ for high-$p_T$ particles [25]. The four considered
models describe the $p_T$ spectra of the identified particles
produced in $pp$ collisions at $\sqrt{s}=2.76$ TeV similarly well
in most of the cases.

\begin{table*}[!htb]
{\scriptsize Table 1. Values of parameters ($T_0$, $\beta_{T}$,
$k$, $p_{0}$, and $n$), normalization constant ($N_{0}$),
$\chi^2$, and degrees of freedom (DOF) corresponding to the fits
of the BGBW model and the inverse power-law [Eqs. (1) and (5)
through Eq. (6) or (7)] in Figs. 1--4 and 8. For better
readability, the collision types, data sources, and collision
energies are listed in the blank spaces of the first two columns.
The results of the very-soft and soft components are listed
together and marked by an asterisk (*) at the end of the line.
\vspace{-.25cm}
\begin{center}
\begin{tabular}{cccccccccc}
\hline\hline Figure & Centrality & Particle & $T_0$ (GeV) & $\beta_{T}$ ($c$) & $k$ & $p_{0}$ (GeV/$c$) & $n$ & $N_0$ & $\chi^2$/DOF \\
\hline
1(a) & 0--20\%   & $\pi^+$   & $0.112\pm0.006$ & $0.43\pm0.01$ & $0.999\pm0.001$ & $5.0\pm0.3$ & $15.9\pm0.4$ & $4.85\pm0.51$ & 37/18\\
$d$-Au & 200 GeV & $K^+$     & $0.128\pm0.008$ & $0.42\pm0.01$ & $0.994\pm0.006$ & $5.8\pm0.3$ & $16.5\pm0.4$ & $0.64\pm0.07$ & 9/15\\
PHENIX &         & $p$       & $0.130\pm0.008$ & $0.39\pm0.01$ & $0.998\pm0.002$ & $5.2\pm0.3$ & $15.2\pm0.4$ & $0.30\pm0.02$ & 64/18\\
1(b) & 0--20\%   & $\pi^-$   & $0.121\pm0.006$ & $0.43\pm0.01$ & $0.999\pm0.001$ & $5.4\pm0.3$ & $16.7\pm0.4$ & $4.30\pm0.50$ & 23/18\\
     &           & $K^-$     & $0.121\pm0.008$ & $0.43\pm0.01$ & $0.995\pm0.004$ & $6.1\pm0.3$ & $17.3\pm0.4$ & $0.60\pm0.06$ & 7/15\\
     &           & $\bar{p}$ & $0.129\pm0.008$ & $0.39\pm0.01$ & $0.999\pm0.001$ & $5.1\pm0.2$ & $16.1\pm0.4$ & $0.24\pm0.02$ & 103/18\\
1(c) & 60--88\%  & $\pi^+$   & $0.104\pm0.006$ & $0.43\pm0.01$ & $0.998\pm0.002$ & $3.5\pm0.2$ & $12.9\pm0.3$ & $1.29\pm0.14$ & 30/18\\
     &           & $K^+$     & $0.116\pm0.008$ & $0.41\pm0.01$ & $0.982\pm0.008$ & $6.4\pm0.2$ & $17.2\pm0.4$ & $0.15\pm0.02$ & 12/15\\
     &           & $p$       & $0.119\pm0.009$ & $0.37\pm0.01$ & $0.996\pm0.004$ & $5.5\pm0.2$ & $15.6\pm0.3$ & $0.07\pm0.01$ & 33/18\\
1(d) & 60--88\%  & $\pi^-$   & $0.104\pm0.006$ & $0.43\pm0.01$ & $0.998\pm0.002$ & $3.5\pm0.2$ & $12.9\pm0.3$ & $1.29\pm0.08$ & 36/18\\
     &           & $K^-$     & $0.115\pm0.008$ & $0.40\pm0.01$ & $0.983\pm0.011$ & $6.0\pm0.2$ & $17.2\pm0.3$ & $0.15\pm0.02$ & 15/15\\
     &           & $\bar{p}$ & $0.119\pm0.008$ & $0.37\pm0.01$ & $0.997\pm0.003$ & $5.5\pm0.2$ & $16.6\pm0.3$ & $0.05\pm0.01$ & 31/18\\
\hline
1(a) & 0--20\%   & $\pi^+$   & $0.111\pm0.006$ & $0.43\pm0.01$ & $0.999\pm0.001$ & $4.4\pm0.2$ & $15.3\pm0.3$ & $9.20\pm0.99$ & 21/18\\
$d$-Au & 200 GeV & $p$       & $0.128\pm0.008$ & $0.37\pm0.01$ & $0.998\pm0.002$ & $5.1\pm0.2$ & $15.9\pm0.3$ & $0.97\pm0.10$ & 18/16\\
1(b) & 0--20\%   & $\pi^-$   & $0.111\pm0.006$ & $0.43\pm0.01$ & $0.999\pm0.001$ & $4.4\pm0.2$ & $15.3\pm0.3$ & $9.2\pm1.00$ & 24/18\\
STAR &           & $\bar{p}$ & $0.127\pm0.005$ & $0.37\pm0.01$ & $0.998\pm0.002$ & $5.1\pm0.1$ & $16.9\pm0.2$ & $0.79\pm0.09$ & 21/16\\
1(c) & 40--100\% & $\pi^+$   & $0.103\pm0.006$ & $0.42\pm0.01$ & $0.999\pm0.001$ & $3.7\pm0.2$ & $13.4\pm0.3$ & $2.78\pm0.28$ & 26/18\\
     &           & $p$       & $0.115\pm0.007$ & $0.37\pm0.01$ & $0.998\pm0.002$ & $6.9\pm0.1$ & $18.2\pm0.3$ & $0.25\pm0.03$ & 33/16\\
1(d) & 40--100\% & $\pi^-$   & $0.103\pm0.006$ & $0.42\pm0.01$ & $0.999\pm0.001$ & $3.7\pm0.2$ & $13.4\pm0.3$ & $2.78\pm0.28$ & 22/18\\
     &           & $\bar{p}$ & $0.112\pm0.006$ & $0.35\pm0.01$ & $0.998\pm0.002$ & $6.4\pm0.1$ & $18.9\pm0.3$ & $0.24\pm0.02$ & 39/16\\
\hline
2(a) &           & $\pi^+$   & $0.104\pm0.006$ & $0.40\pm0.01$ & $0.999\pm0.001$ & $2.2\pm0.1$ & $11.2\pm0.3$ & $0.64\pm0.07$ & 22/23\\
$pp$ & 200 GeV   & $K^+$     & $0.114\pm0.008$ & $0.41\pm0.01$ & $0.999\pm0.001$ & $3.0\pm0.1$ & $12.4\pm0.3$ & $0.07\pm0.01$ & 8/18\\
STAR &           & $p$       & $0.116\pm0.008$ & $0.34\pm0.01$ & $0.999\pm0.001$ & $3.1\pm0.2$ & $12.6\pm0.3$ & $0.05\pm0.01$ & 29/22\\
2(b) &           & $\pi^-$   & $0.104\pm0.006$ & $0.40\pm0.01$ & $0.999\pm0.001$ & $2.2\pm0.1$ & $11.3\pm0.3$ & $0.64\pm0.07$ & 27/23\\
     &           & $K^-$     & $0.114\pm0.008$ & $0.41\pm0.01$ & $0.999\pm0.001$ & $3.2\pm0.1$ & $13.5\pm0.3$ & $0.07\pm0.01$ & 4/18\\
     &           & $\bar{p}$ & $0.116\pm0.008$ & $0.34\pm0.01$ & $0.998\pm0.002$ & $3.1\pm0.2$ & $13.7\pm0.4$ & $0.04\pm0.01$ & 46/22\\
\hline
3(a) & 0--5\%    & $\pi^\pm$     & $0.136\pm0.008$ & $0.43\pm0.01$ & $0.999\pm0.001$ & $2.1\pm0.1$ & $7.6\pm0.3$ & $18.70\pm1.99$& 320/49*\\
$p$-Pb & 5.02 TeV& $K^\pm$       & $0.193\pm0.009$ & $0.43\pm0.01$ & $0.997\pm0.003$ & $2.7\pm0.1$ & $7.3\pm0.3$ & $2.84\pm0.41$ & 71/45\\
ALICE &          & $p$+$\bar{p}$ & $0.195\pm0.009$ & $0.42\pm0.01$ & $0.999\pm0.001$ & $3.5\pm0.2$ & $8.8\pm0.3$ & $1.10\pm0.11$ & 172/43\\
3(b) & 80--100\% & $\pi^\pm$     & $0.112\pm0.008$ & $0.43\pm0.01$ & $0.988\pm0.006$ & $1.3\pm0.1$ & $7.4\pm0.3$ & $1.91\pm0.20$ & 234/52\\
     &           & $K^\pm$       & $0.139\pm0.008$ & $0.41\pm0.01$ & $0.990\pm0.006$ & $3.3\pm0.1$ & $8.9\pm0.3$ & $0.25\pm0.02$ & 119/45\\
     &           & $p$+$\bar{p}$ & $0.156\pm0.009$ & $0.37\pm0.01$ & $0.993\pm0.006$ & $3.9\pm0.1$ & $10.1\pm0.3$ & $0.10\pm0.01$& 225/43\\
\hline
4    &           & $\pi^\pm$     & $0.111\pm0.008$ & $0.43\pm0.01$ & $0.994\pm0.005$ & $1.9\pm0.1$ & $8.1\pm0.3$ & $3.60\pm0.35$ & 382/57\\
$pp$ &  2.76 TeV & $K^\pm$       & $0.143\pm0.008$ & $0.42\pm0.01$ & $0.990\pm0.005$ & $2.9\pm0.1$ & $8.6\pm0.3$ & $0.45\pm0.05$ & 119/52\\
ALICE &          & $p$+$\bar{p}$ & $0.152\pm0.009$ & $0.36\pm0.01$ & $0.991\pm0.005$ & $2.6\pm0.1$ & $9.5\pm0.3$ & $0.19\pm0.01$ & 214/43\\
\hline
8(a)  & 0--20\%  & $\pi^\pm$     & $0.107\pm0.006$ & $0.41\pm0.01$ & $0.999\pm0.001$ & $4.4\pm0.3$ & $14.5\pm0.4$ & $103.61\pm11.37$& 28/23\\
Cu-Cu & 200 GeV  & $K^\pm$       & $0.122\pm0.011$ & $0.41\pm0.02$ & $0.997\pm0.003$ & $6.1\pm0.3$ & $16.3\pm0.4$ & $12.52\pm1.26$  & 1/10\\
      &          & $p$+$\bar{p}$ & $0.125\pm0.008$ & $0.38\pm0.01$ & $0.999\pm0.001$ & $5.2\pm0.3$ & $15.7\pm0.4$ & $7.85\pm0.77$   & 5/21\\
8(b)  & 40--94\% & $\pi^\pm$     & $0.101\pm0.005$ & $0.43\pm0.01$ & $0.999\pm0.001$ & $4.3\pm0.2$ & $14.5\pm0.3$ & $8.29\pm0.81$   & 18/23\\
      &          & $K^\pm$       & $0.111\pm0.008$ & $0.40\pm0.01$ & $0.996\pm0.003$ & $5.9\pm0.2$ & $16.9\pm0.3$ & $1.28\pm0.11$   & 1/10\\
      &          & $p$+$\bar{p}$ & $0.114\pm0.009$ & $0.37\pm0.01$ & $0.996\pm0.003$ & $6.4\pm0.2$ & $19.9\pm0.2$ & $0.50\pm0.05$   & 15/21\\
\hline
\end{tabular}
\end{center}}
\end{table*}

\begin{figure*}[!htb]
\begin{center}
\includegraphics[width=7.5cm]{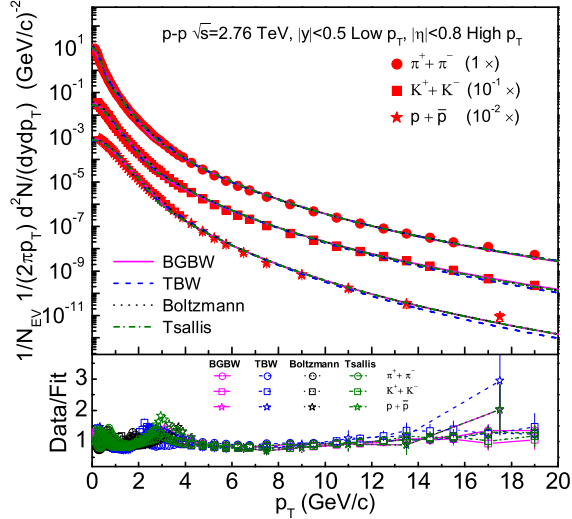}
\end{center} \vskip-.5cm
{\caption{\small (Color online) Spectra of $\pi^++\pi^-$,
$K^++K^-$, and $p+\bar p$ produced in $pp$ collisions at
$\sqrt{s}=2.76$ TeV. The symbols represent the experimental data
of the ALICE collaboration measured in $|y|<0.5$ for low-$p_T$
particles and in $|\eta|<0.8$ for high-$p_T$ particles [25].}}
\end{figure*}

\begin{table*}[!htb]
{\scriptsize Table 2. Values of parameters ($T_0$, $q$,
$\beta_{T}$, $k$, $\ p_{0}$, and $n$), normalization constant
($N_{0}$), $\chi^2$, and DOF corresponding to the fits of the TBW
model and the inverse power-law [Eqs. (2) and (5) through Eq. (6)
or (7)] in Figs. 1--4 and 8, where the columns of centrality and
particle are the same as those in Table 1; thus, these are
omitted.
\begin{center}
\begin{tabular}{ccccccccc}
\hline\hline  Figure & $T_0$ (GeV) & $q$ & $\beta_{T}$ ($c$) & $k$ & $p_{0}$ (GeV/$c$) & $n$ & $N_0$ & $\chi^2$/DOF \\
\hline
1(a)  & $0.108\pm0.006$ & $1.025\pm0.007$ & $0.46\pm0.01$ & $0.991\pm0.005$ & $4.8\pm0.3$ & $16.2\pm0.4$ & $3.86\pm0.39$ & 46/17\\
$d$-Au& $0.118\pm0.008$ & $1.026\pm0.008$ & $0.46\pm0.01$ & $0.981\pm0.006$ & $5.9\pm0.3$ & $16.0\pm0.4$ & $0.57\pm0.06$ & 24/14\\
PHENIX& $0.119\pm0.008$ & $1.018\pm0.007$ & $0.45\pm0.01$ & $0.996\pm0.004$ & $5.1\pm0.2$ & $15.9\pm0.4$ & $0.25\pm0.02$ & 19/17\\
1(b)  & $0.108\pm0.006$ & $1.025\pm0.007$ & $0.46\pm0.01$ & $0.992\pm0.005$ & $4.8\pm0.3$ & $16.4\pm0.4$ & $3.86\pm0.39$ & 56/17\\
      & $0.118\pm0.008$ & $1.026\pm0.008$ & $0.46\pm0.01$ & $0.983\pm0.009$ & $5.9\pm0.3$ & $17.0\pm0.4$ & $0.57\pm0.06$ & 34/14\\
      & $0.118\pm0.008$ & $1.018\pm0.007$ & $0.45\pm0.01$ & $0.996\pm0.004$ & $5.1\pm0.2$ & $16.3\pm0.4$ & $0.20\pm0.02$ & 36/17\\
1(c)  & $0.088\pm0.006$ & $1.045\pm0.008$ & $0.46\pm0.01$ & $0.994\pm0.004$ & $3.5\pm0.2$ & $13.7\pm0.3$ & $1.02\pm0.10$ & 34/17\\
      & $0.090\pm0.008$ & $1.029\pm0.008$ & $0.46\pm0.01$ & $0.955\pm0.011$ & $6.4\pm0.3$ & $17.5\pm0.4$ & $0.13\pm0.01$ & 9/14\\
      & $0.098\pm0.008$ & $1.012\pm0.007$ & $0.44\pm0.01$ & $0.990\pm0.006$ & $5.5\pm0.2$ & $15.9\pm0.2$ & $0.06\pm0.01$ & 37/17\\
1(d)  & $0.088\pm0.006$ & $1.045\pm0.008$ & $0.46\pm0.01$ & $0.994\pm0.006$ & $3.5\pm0.2$ & $13.7\pm0.3$ & $1.02\pm0.10$ & 46/17\\
      & $0.090\pm0.008$ & $1.029\pm0.008$ & $0.46\pm0.01$ & $0.957\pm0.011$ & $6.7\pm0.3$ & $18.6\pm0.4$ & $0.13\pm0.01$ & 11/14\\
      & $0.097\pm0.008$ & $1.012\pm0.007$ & $0.43\pm0.01$ & $0.992\pm0.006$ & $5.5\pm0.2$ & $17.1\pm0.3$ & $0.05\pm0.01$ & 48/17\\
\hline
1(a)  & $0.106\pm0.006$ & $1.020\pm0.008$ & $0.46\pm0.01$ & $0.995\pm0.004$ & $4.4\pm0.2$ & $15.6\pm0.4$ & $7.07\pm0.76$ & 38/17\\
$d$-Au& $0.115\pm0.008$ & $1.010\pm0.007$ & $0.38\pm0.01$ & $0.998\pm0.002$ & $4.4\pm0.2$ & $15.6\pm0.4$ & $0.96\pm0.10$ & 35/11*\\
1(b)  & $0.106\pm0.006$ & $1.020\pm0.008$ & $0.46\pm0.01$ & $0.995\pm0.004$ & $4.4\pm0.2$ & $15.6\pm0.4$ & $7.07\pm0.76$ & 39/17\\
STAR  & $0.116\pm0.008$ & $1.008\pm0.005$ & $0.44\pm0.01$ & $0.997\pm0.003$ & $5.1\pm0.2$ & $17.7\pm0.4$ & $0.73\pm0.07$ & 44/15\\
1(c)  & $0.085\pm0.006$ & $1.038\pm0.008$ & $0.46\pm0.01$ & $0.996\pm0.004$ & $3.7\pm0.2$ & $13.9\pm0.3$ & $2.50\pm0.25$ & 33/17\\
      & $0.090\pm0.008$ & $1.008\pm0.007$ & $0.35\pm0.01$ & $0.998\pm0.002$ & $6.9\pm0.2$ & $19.6\pm0.4$ & $0.31\pm0.02$ & 29/11*\\
1(d)  & $0.085\pm0.006$ & $1.038\pm0.008$ & $0.46\pm0.01$ & $0.996\pm0.004$ & $3.7\pm0.2$ & $13.9\pm0.3$ & $2.54\pm0.25$ & 48/17\\
      & $0.094\pm0.008$ & $1.016\pm0.007$ & $0.44\pm0.01$ & $0.996\pm0.004$ & $5.9\pm0.2$ & $19.4\pm0.3$ & $0.19\pm0.02$ & 53/15\\
\hline
2(a)  & $0.089\pm0.006$ & $1.023\pm0.008$ & $0.44\pm0.01$ & $0.997\pm0.003$ & $2.2\pm0.1$ & $11.2\pm0.4$ & $0.62\pm0.06$ & 41/22\\
$pp$  & $0.098\pm0.008$ & $1.029\pm0.009$ & $0.43\pm0.01$ & $0.996\pm0.004$ & $3.0\pm0.2$ & $12.8\pm0.4$ & $0.07\pm0.01$ & 29/17\\
STAR  & $0.104\pm0.009$ & $1.006\pm0.001$ & $0.39\pm0.01$ & $0.996\pm0.004$ & $3.1\pm0.2$ & $13.5\pm0.4$ & $0.05\pm0.01$ & 55/21\\
2(b)  & $0.089\pm0.006$ & $1.023\pm0.008$ & $0.44\pm0.01$ & $0.997\pm0.003$ & $2.2\pm0.1$ & $11.5\pm0.4$ & $0.62\pm0.06$ & 52/22\\
      & $0.098\pm0.008$ & $1.029\pm0.009$ & $0.43\pm0.01$ & $0.996\pm0.004$ & $3.0\pm0.2$ & $13.8\pm0.4$ & $0.07\pm0.01$ & 26/17\\
      & $0.104\pm0.009$ & $1.006\pm0.001$ & $0.39\pm0.01$ & $0.996\pm0.004$ & $3.1\pm0.2$ & $13.9\pm0.4$ & $0.04\pm0.01$ & 84/21\\
\hline
3(a)  & $0.107\pm0.007$ & $1.001\pm0.001$ & $0.48\pm0.01$ & $0.999\pm0.001$ & $2.2\pm0.1$ & $7.7\pm0.3$ & $20.98\pm2.00$& 323/47*\\
$p$-Pb& $0.188\pm0.009$ & $1.012\pm0.006$ & $0.48\pm0.01$ & $0.995\pm0.004$ & $2.7\pm0.2$ & $7.8\pm0.3$ & $2.78\pm0.29$ & 436/44\\
ALICE & $0.198\pm0.009$ & $1.013\pm0.008$ & $0.47\pm0.01$ & $0.999\pm0.001$ & $3.5\pm0.2$ & $9.1\pm0.3$ & $1.10\pm0.10$ & 223/42\\
3(b)  & $0.089\pm0.006$ & $1.001\pm0.001$ & $0.45\pm0.01$ & $0.999\pm0.001$ & $1.4\pm0.1$ & $7.3\pm0.3$ & $2.21\pm0.20$ & 606/43*\\
      & $0.113\pm0.008$ & $1.023\pm0.006$ & $0.45\pm0.01$ & $0.976\pm0.010$ & $3.3\pm0.2$ & $9.1\pm0.3$ & $0.23\pm0.02$ & 325/44\\
      & $0.115\pm0.009$ & $1.002\pm0.001$ & $0.45\pm0.01$ & $0.982\pm0.010$ & $3.9\pm0.2$ & $10.6\pm0.3$ & $0.09\pm0.01$& 493/42\\
\hline
4     & $0.089\pm0.006$ & $1.001\pm0.001$ & $0.45\pm0.01$ & $0.999\pm0.001$ & $1.7\pm0.1$ & $7.8\pm0.3$ & $4.00\pm0.31$ & 485/48*\\
$pp$  & $0.113\pm0.008$ & $1.013\pm0.006$ & $0.48\pm0.01$ & $0.975\pm0.010$ & $2.9\pm0.1$ & $9.0\pm0.3$ & $0.46\pm0.05$ & 376/51\\
ALICE & $0.116\pm0.008$ & $1.004\pm0.001$ & $0.44\pm0.01$ & $0.975\pm0.010$ & $2.5\pm0.2$ & $9.9\pm0.3$ & $0.20\pm0.02$ & 494/42\\
\hline
8(a)  & $0.101\pm0.007$ & $1.027\pm0.009$ & $0.47\pm0.02$ & $0.999\pm0.001$ & $4.4\pm0.2$ & $14.8\pm0.4$ & $66.17\pm7.10$& 27/22\\
Cu-Cu & $0.110\pm0.008$ & $1.026\pm0.008$ & $0.46\pm0.02$ & $0.996\pm0.004$ & $6.2\pm0.3$ & $16.5\pm0.4$ & $12.23\pm1.20$& 3/9\\
      & $0.114\pm0.008$ & $1.020\pm0.007$ & $0.45\pm0.01$ & $0.999\pm0.001$ & $5.2\pm0.3$ & $16.3\pm0.4$ & $6.02\pm0.60$ & 4/20\\
8(b)  & $0.085\pm0.007$ & $1.052\pm0.008$ & $0.47\pm0.02$ & $0.999\pm0.001$ & $4.3\pm0.2$ & $14.7\pm0.3$ & $6.58\pm0.68$ & 19/22\\
      & $0.090\pm0.008$ & $1.029\pm0.008$ & $0.47\pm0.02$ & $0.996\pm0.004$ & $6.0\pm0.3$ & $16.7\pm0.3$ & $1.08\pm0.01$ & 3/9\\
      & $0.095\pm0.008$ & $1.012\pm0.008$ & $0.46\pm0.01$ & $0.992\pm0.004$ & $6.6\pm0.2$ & $21.3\pm0.4$ & $0.38\pm0.04$ & 12/20\\
\hline
\end{tabular}
\end{center}}
\end{table*}

\begin{table*}[!htb]
{\scriptsize Table 3. Values of parameters ($T$, $k$, $p_{0}$, and
$n$ ), normalization constant ($N_{0}$), $\chi^2$, and DOF
corresponding to the fits of the Boltzmann distribution and the
inverse power-law [Eqs. (3) and (5) through Eq. (6) or (7)] in
Figs. 1--4 and 8.
\begin{center}
\begin{tabular}{ccccccccc}
\hline\hline  Figure & Centrality & Particle & $T$ (GeV) & $k$ & $p_{0}$ (GeV/$c$) & $n$ & $N_0$ & $\chi^2$/DOF \\
\hline
1(a) &  0--20\%  & $\pi^+$   & $0.179\pm0.006$ & $0.992\pm0.005$ & $4.9\pm0.2$ & $16.8\pm0.3$ & $3.70\pm0.35$ & 28/17* \\
$d$-Au & 200 GeV & $K^+$     & $0.243\pm0.009$ & $0.976\pm0.011$ & $5.9\pm0.2$ & $16.9\pm0.3$ & $0.60\pm0.05$ & 39/16 \\
PHENIX &         & $p$       & $0.293\pm0.009$ & $0.991\pm0.006$ & $5.1\pm0.2$ & $15.8\pm0.3$ & $0.25\pm0.02$ & 24/19 \\
1(b) &  0--20\%  &$\pi^-$    & $0.179\pm0.006$ & $0.993\pm0.006$ & $4.8\pm0.2$ & $16.8\pm0.3$ & $3.70\pm0.35$ & 29/17* \\
     &           &$K^-$      & $0.240\pm0.009$ & $0.974\pm0.011$ & $5.6\pm0.2$ & $16.9\pm0.3$ & $0.58\pm0.05$ & 37/16 \\
     &           & $\bar{p}$ & $0.290\pm0.009$ & $0.993\pm0.005$ & $5.0\pm0.2$ & $16.5\pm0.3$ & $0.20\pm0.02$ & 30/19 \\
1(c) & 60--88\%  & $\pi^+$   & $0.148\pm0.006$ & $0.995\pm0.005$ & $3.5\pm0.1$ & $13.5\pm0.2$ & $1.14\pm0.01$ & 61/17* \\
     &           & $K^+$     & $0.200\pm0.009$ & $0.950\pm0.011$ & $6.4\pm0.3$ & $17.7\pm0.3$ & $0.15\pm0.01$ & 18/16 \\
     &           & $p$       & $0.247\pm0.009$ & $0.993\pm0.005$ & $5.3\pm0.2$ & $15.5\pm0.3$ & $0.07\pm0.01$ & 42/19 \\
1(d) & 60--88\%  & $\pi^-$   & $0.148\pm0.006$ & $0.995\pm0.004$ & $3.5\pm0.1$ & $13.5\pm0.2$ & $1.14\pm0.01$ & 70/17*\\
     &           & $K^-$     & $0.200\pm0.009$ & $0.954\pm0.012$ & $6.2\pm0.3$ & $17.9\pm0.3$ & $0.14\pm0.01$ & 17/16\\
     &           & $\bar{p}$ & $0.247\pm0.009$ & $0.993\pm0.005$ & $5.0\pm0.2$ & $16.5\pm0.3$ & $0.05\pm0.01$ & 28/19 \\
\hline
1(a) &  0--20\%  & $\pi^+$   & $0.172\pm0.007$ & $0.999\pm0.001$ & $4.1\pm0.1$ & $15.0\pm0.3$ & $7.70\pm0.70$ & 42/17* \\
$d$-Au & 200 GeV & $p$       & $0.208\pm0.009$ & $0.999\pm0.001$ & $5.8\pm0.2$ & $16.5\pm0.3$ & $1.07\pm0.10$ & 28/15* \\
1(b) &  0--20\%  & $\pi^-$   & $0.172\pm0.007$ & $0.999\pm0.001$ & $4.1\pm0.1$ & $15.0\pm0.3$ & $7.70\pm0.70$ & 36/17* \\
STAR &           & $\bar{p}$ & $0.253\pm0.008$ & $0.997\pm0.003$ & $5.2\pm0.2$ & $17.2\pm0.3$ & $0.73\pm0.06$ & 33/17 \\
1(c) & 40--100\% & $\pi^+$   & $0.143\pm0.007$ & $0.998\pm0.002$ & $3.1\pm0.1$ & $12.4\pm0.2$ & $2.63\pm0.24$ & 59/17* \\
     &           & $p$       & $0.219\pm0.009$ & $0.991\pm0.005$ & $5.8\pm0.2$ & $17.8\pm0.3$ & $0.27\pm0.02$ & 37/17 \\
1(d) & 40--100\% & $\pi^-$   & $0.143\pm0.007$ & $0.998\pm0.002$ & $3.1\pm0.1$ & $12.4\pm0.2$ & $2.63\pm0.24$ & 49/17* \\
     &           & $\bar{p}$ & $0.217\pm0.009$ & $0.992\pm0.005$ & $5.4\pm0.1$ & $18.5\pm0.3$ & $0.22\pm0.02$ & 30/17 \\
\hline
2(a) &           & $\pi^+$   & $0.144\pm0.007$ & $0.999\pm0.001$ & $2.0\pm0.1$ & $10.8\pm0.3$ & $0.55\pm0.33$ & 35/22* \\
$pp$ & 200 GeV   & $K^+$     & $0.203\pm0.009$ & $0.989\pm0.007$ & $3.3\pm0.1$ & $13.3\pm0.3$ & $0.07\pm0.01$ & 31/19 \\
STAR &           & $p$       & $0.234\pm0.009$ & $0.997\pm0.003$ & $3.2\pm0.2$ & $13.6\pm0.3$ & $0.05\pm0.01$ & 121/23 \\
2(b) &           & $\pi^-$   & $0.144\pm0.007$ & $0.999\pm0.001$ & $2.0\pm0.1$ & $10.8\pm0.3$ & $0.55\pm0.33$ & 47/22* \\
     &           & $K^-$     & $0.203\pm0.009$ & $0.991\pm0.005$ & $3.1\pm0.1$ & $13.7\pm0.3$ & $0.07\pm0.01$ & 21/19 \\
     &           & $\bar{p}$ & $0.230\pm0.009$ & $0.996\pm0.004$ & $3.1\pm0.2$ & $14.3\pm0.3$ & $0.04\pm0.01$ & 91/23 \\
\hline
3(a) &  0--5\%   & $\pi^\pm$    & $0.163\pm0.008$ & $0.999\pm0.001$ & $2.0\pm0.1$ & $7.7\pm0.3$ & $22.14\pm2.10$& 852/49* \\
$p$-Pb & 5.02 TeV& $K^\pm$      & $0.297\pm0.008$ & $0.992\pm0.005$ & $3.4\pm0.1$ & $8.6\pm0.3$ & $2.87\pm0.29$ & 110/44* \\
ALICE &          & $p$+$\bar{p}$& $0.381\pm0.009$ & $0.997\pm0.003$ & $3.2\pm0.1$ & $9.4\pm0.3$ & $1.17\pm0.01$ & 138/42* \\
3(b) & 80--100\% & $\pi^\pm$    & $0.123\pm0.009$ & $0.999\pm0.001$ & $1.4\pm0.1$ & $7.2\pm0.3$ & $2.23\pm0.19$ & 935/49* \\
     &           & $K^\pm$      & $0.212\pm0.010$ & $0.995\pm0.005$ & $3.8\pm0.1$ & $9.0\pm0.3$ & $0.28\pm0.02$ & 403/44* \\
     &           & $p$+$\bar{p}$& $0.235\pm0.010$ & $0.997\pm0.003$ & $3.3\pm0.1$ & $9.8\pm0.3$ & $0.11\pm0.01$ & 128/42* \\
\hline
4    &           & $\pi^\pm$    & $0.123\pm0.008$ & $0.997\pm0.003$ & $1.7\pm0.1$ & $7.9\pm0.3$ & $4.14\pm0.31$ & 688/54* \\
$pp$ & 2.76 TeV  & $K^\pm$      & $0.205\pm0.009$ & $0.988\pm0.010$ & $2.8\pm0.1$ & $8.7\pm0.3$ & $0.49\pm0.06$ & 178/51* \\
ALICE &          & $p$+$\bar{p}$& $0.241\pm0.009$ & $0.995\pm0.005$ & $2.6\pm0.1$ & $9.5\pm0.3$ & $0.20\pm0.02$ & 104/42* \\
\hline
8(a)  & 0--20\%  & $\pi^\pm$    & $0.179\pm0.008$ & $0.999\pm0.001$ & $4.2\pm0.1$ & $14.5\pm0.2$ & $72.07\pm7.00$ & 35/22* \\
Cu-Cu & 200 GeV  & $K^\pm$      & $0.231\pm0.010$ & $0.991\pm0.006$ & $5.9\pm0.2$ & $16.9\pm0.3$ & $12.37\pm1.21$ & 5/11 \\
      &          &$p$+$\bar{p}$ & $0.296\pm0.009$ & $0.999\pm0.001$ & $5.4\pm0.2$ & $17.0\pm0.4$ & $5.81\pm0.67$  & 6/22 \\
8(b)  & 40--94\% & $\pi^\pm$    & $0.139\pm0.006$ & $0.999\pm0.001$ & $3.8\pm0.1$ & $13.9\pm0.2$ & $10.95\pm0.89$ & 23/22* \\
      &          & $K^\pm$      & $0.179\pm0.009$ & $0.994\pm0.005$ & $4.9\pm0.1$ & $16.2\pm0.3$ & $1.16\pm0.13$  & 1/9* \\
      &          &$p$+$\bar{p}$ & $0.245\pm0.008$ & $0.999\pm0.001$ & $4.8\pm0.2$ & $17.7\pm0.3$ & $0.43\pm0.06$  & 3/20* \\
\hline
\end{tabular}
\end{center}}
\end{table*}

\begin{table*}[!htb]
{\scriptsize Table 4. Values of parameters ($T$, $q$, $k$,
$p_{0}$, and $n$), normalization constant ($N_{0}$), $\chi^2$, and
DOF corresponding to the fits of the Tsallis distribution and the
inverse power-law [Eqs. (4) and (5) through Eq. (6) or (7)] in
Figs. 1--4 and 8. \vspace{-.25cm}
\begin{center}
\begin{tabular}{cccccccccc}
\hline\hline  Figure & Centrality & Particle & $T$ (GeV) & $q$ & $k$ & $p_{0}$ (GeV/$c$) & $n$ & $N_0$ & $\chi^2$/DOF \\
\hline
1(a) &  0--20\%  & $\pi^+$   & $0.134\pm0.008$ & $1.082\pm0.009$ & $0.994\pm0.005$ & $4.8\pm0.2$ & $16.3\pm0.4$ & $3.93\pm0.36$ & 32/18 \\
$d$-Au & 200 GeV & $K^+$     & $0.189\pm0.009$ & $1.052\pm0.010$ & $0.980\pm0.010$ & $6.1\pm0.2$ & $16.9\pm0.4$ & $0.57\pm0.06$ & 13/15 \\
PHENIX &         & $p$       & $0.272\pm0.009$ & $1.015\pm0.007$ & $0.999\pm0.001$ & $5.5\pm0.2$ & $15.3\pm0.4$ & $0.28\pm0.02$ & 49/18 \\
1(b) &  0--20\%  &$\pi^-$    & $0.134\pm0.008$ & $1.082\pm0.009$ & $0.993\pm0.005$ & $4.8\pm0.2$ & $16.8\pm0.4$ & $3.68\pm0.36$ & 33/18 \\
     &           &$K^-$      & $0.189\pm0.009$ & $1.052\pm0.010$ & $0.982\pm0.011$ & $6.0\pm0.2$ & $17.1\pm0.4$ & $0.56\pm0.06$ & 18/15 \\
     &           & $\bar{p}$ & $0.273\pm0.009$ & $1.015\pm0.007$ & $0.999\pm0.001$ & $5.4\pm0.2$ & $16.0\pm0.4$ & $0.18\pm0.02$ & 44/18 \\
1(c) & 60--88\%  & $\pi^+$   & $0.108\pm0.008$ & $1.099\pm0.009$ & $0.999\pm0.001$ & $3.6\pm0.1$ & $13.1\pm0.3$ & $1.20\pm0.12$ & 37/18 \\
     &           & $K^+$     & $0.141\pm0.009$ & $1.083\pm0.011$ & $0.984\pm0.012$ & $6.6\pm0.2$ & $17.5\pm0.4$ & $0.15\pm0.02$ & 16/15 \\
     &           & $p$       & $0.194\pm0.010$ & $1.035\pm0.010$ & $0.996\pm0.004$ & $5.8\pm0.2$ & $15.9\pm0.4$ & $0.07\pm0.01$ & 22/18 \\
1(d) & 60--88\%  & $\pi^-$   & $0.108\pm0.008$ & $1.099\pm0.009$ & $0.999\pm0.001$ & $3.5\pm0.1$ & $12.9\pm0.3$ & $1.17\pm0.12$ & 36/18 \\
     &           & $K^-$     & $0.141\pm0.009$ & $1.083\pm0.011$ & $0.979\pm0.012$ & $6.5\pm0.2$ & $17.8\pm0.4$ & $0.13\pm0.02$ & 11/15 \\
     &           & $\bar{p}$ & $0.194\pm0.010$ & $1.035\pm0.010$ & $0.998\pm0.002$ & $5.8\pm0.2$ & $16.6\pm0.4$ & $0.05\pm0.01$ & 30/18 \\
\hline
1(a) &  0--20\%  & $\pi^+$   & $0.129\pm0.008$ & $1.076\pm0.009$ & $0.997\pm0.003$ & $4.4\pm0.1$ & $15.4\pm0.4$ & $8.03\pm0.80$ & 26/18 \\
$d$-Au & 200 GeV & $p$       & $0.221\pm0.009$ & $1.005\pm0.005$ & $0.999\pm0.001$ & $5.9\pm0.2$ & $16.8\pm0.3$ & $1.01\pm0.09$ & 23/13* \\
1(b) &  0--20\%  &$\pi^-$    & $0.129\pm0.008$ & $1.076\pm0.009$ & $0.997\pm0.003$ & $4.4\pm0.1$ & $15.4\pm0.4$ & $8.03\pm0.80$ & 27/18 \\
STAR &           & $\bar{p}$ & $0.260\pm0.009$ & $1.009\pm0.005$ & $0.999\pm0.001$ & $5.7\pm0.2$ & $17.3\pm0.3$ & $0.68\pm0.07$ & 46/16 \\
1(c) & 40--100\% & $\pi^+$   & $0.104\pm0.008$ & $1.089\pm0.009$ & $0.998\pm0.002$ & $3.4\pm0.1$ & $13.1\pm0.3$ & $2.62\pm0.25$ & 32/18 \\
     &           & $p$       & $0.173\pm0.009$ & $1.011\pm0.005$ & $0.999\pm0.001$ & $6.3\pm0.2$ & $17.0\pm0.3$ & $0.31\pm0.03$ & 33/13* \\
1(d) & 40--100\% & $\pi^-$   & $0.104\pm0.008$ & $1.089\pm0.009$ & $0.998\pm0.002$ & $3.4\pm0.1$ & $13.1\pm0.3$ & $2.62\pm0.25$ & 26/18 \\
     &           & $\bar{p}$ & $0.189\pm0.009$ & $1.036\pm0.005$ & $0.999\pm0.001$ & $5.4\pm0.2$ & $17.7\pm0.3$ & $0.19\pm0.02$ & 35/16 \\
\hline
2(a) &           & $\pi^+$   & $0.120\pm0.008$ & $1.051\pm0.009$ & $0.997\pm0.003$ & $2.1\pm0.1$ & $10.9\pm0.3$ & $0.60\pm0.05$ & 44/23 \\
$pp$ & 200 GeV   & $K^+$     & $0.153\pm0.009$ & $1.057\pm0.011$ & $0.997\pm0.003$ & $3.5\pm0.1$ & $13.2\pm0.3$ & $0.07\pm0.01$ & 14/18 \\
STAR &           & $p$       & $0.190\pm0.009$ & $1.019\pm0.009$ & $0.997\pm0.003$ & $3.3\pm0.1$ & $13.3\pm0.4$ & $0.05\pm0.01$ & 32/22 \\
2(b) &           & $\pi^-$   & $0.120\pm0.008$ & $1.056\pm0.009$ & $0.997\pm0.003$ & $2.1\pm0.1$ & $11.0\pm0.3$ & $0.56\pm0.05$ & 45/23 \\
     &           & $K^-$     & $0.153\pm0.009$ & $1.057\pm0.011$ & $0.998\pm0.002$ & $3.5\pm0.1$ & $13.9\pm0.3$ & $0.07\pm0.01$ & 7/18 \\
     &           & $\bar{p}$ & $0.190\pm0.009$ & $1.019\pm0.009$ & $0.997\pm0.003$ & $3.3\pm0.1$ & $13.9\pm0.4$ & $0.04\pm0.01$ & 42/22 \\
\hline
3(a) &  0--5\%   & $\pi^\pm$     & $0.156\pm0.008$ & $1.031\pm0.012$ & $0.999\pm0.001$ & $2.2\pm0.1$ & $7.7\pm0.3$ & $21.20\pm1.91$& 934/45* \\
$p$-Pb & 5.02 TeV& $K^\pm$       & $0.262\pm0.008$ & $1.059\pm0.011$ & $0.995\pm0.005$ & $3.0\pm0.1$ & $7.8\pm0.3$ & $2.78\pm0.28$ & 261/45 \\
ALICE &          & $p$+$\bar{p}$ & $0.351\pm0.009$ & $1.035\pm0.009$ & $0.999\pm0.001$ & $3.4\pm0.1$ & $9.1\pm0.3$ & $1.09\pm0.01$ & 97/43 \\
3(b) & 80--100\% & $\pi^\pm$     & $0.111\pm0.008$ & $1.042\pm0.009$ & $0.998\pm0.002$ & $1.4\pm0.1$ & $7.3\pm0.3$ & $2.15\pm0.20$ & 389/49* \\
     &           & $K^\pm$       & $0.171\pm0.008$ & $1.068\pm0.012$ & $0.986\pm0.010$ & $3.8\pm0.1$ & $9.3\pm0.3$ & $0.23\pm0.02$ & 282/45 \\
     &           & $p$+$\bar{p}$ & $0.192\pm0.009$ & $1.056\pm0.011$ & $0.993\pm0.005$ & $3.4\pm0.1$ & $9.8\pm0.3$ & $0.10\pm0.01$ & 230/43 \\
\hline
4    &           & $\pi^\pm$     & $0.112\pm0.008$ & $1.042\pm0.004$ & $0.997\pm0.003$ & $1.7\pm0.1$ & $7.9\pm0.3$ & $3.90\pm0.36$ & 461/54* \\
$pp$ & 2.76 TeV  & $K^\pm$       & $0.175\pm0.009$ & $1.071\pm0.011$ & $0.985\pm0.010$ & $2.8\pm0.1$ & $8.7\pm0.3$ & $0.44\pm0.06$ & 253/52 \\
ALICE &          & $p$+$\bar{p}$ & $0.223\pm0.009$ & $1.029\pm0.008$ & $0.988\pm0.010$ & $2.6\pm0.1$ & $9.5\pm0.3$ & $0.19\pm0.02$ & 373/43 \\
\hline
8(a)  & 0--20\%  & $\pi^+$ & $0.131\pm0.007$ & $1.070\pm0.006$ & $0.999\pm0.001$ & $4.4\pm0.2$ & $14.6\pm0.3$ & $73.32\pm8.01$ & 23/23 \\
Cu-Cu & 200 GeV  & $K^+$   & $0.173\pm0.011$ & $1.055\pm0.010$ & $0.997\pm0.003$ & $6.4\pm0.2$ & $16.9\pm0.4$ & $12.42\pm1.31$ & 1/10 \\
      &          & $p$     & $0.250\pm0.009$ & $1.018\pm0.006$ & $0.995\pm0.005$ & $5.4\pm0.2$ & $16.4\pm0.3$ & $6.30\pm0.75$  & 6/21 \\
8(b)  & 40--94\% & $\pi^+$ & $0.105\pm0.006$ & $1.096\pm0.006$ & $0.999\pm0.001$ & $4.4\pm0.3$ & $14.1\pm0.4$ & $7.90\pm0.83$  & 23/23 \\
      &          & $K^+$   & $0.139\pm0.009$ & $1.076\pm0.009$ & $0.998\pm0.002$ & $5.2\pm0.3$ & $15.9\pm0.4$ & $1.08\pm0.12$  & 1/10 \\
      &          & $p$     & $0.197\pm0.009$ & $1.042\pm0.006$ & $0.995\pm0.005$ & $5.1\pm0.2$ & $17.9\pm0.3$ & $0.43\pm0.04$  & 9/21 \\
\hline
\end{tabular}
\end{center}}
\end{table*}

It should be noted that although we have used several free
parameters in each fit, these parameters are restricted and
irrelevant. A small number of them (1--3 parameters) are sensitive
to the very-soft component which describes the very-low $p_T$
range from 0 to 0.5--1.5 GeV/$c$ in certain cases. The same number
of parameters (1--3) are sensitive to the soft component
describing the low $p_T$ range from 0.5--1.5 GeV/$c$ to 2.5--3.5
GeV/$c$ in certain cases or typically in the range from 0 to
2.5--3.5 GeV/$c$. While the final two parameters ($p_0$ and $n$)
are sensitive to the hard component describing the wide $p_T$
range from 2.5--3.5 GeV/$c$ to the maximum. In certain cases, the
data in the very-low $p_T$ range is not available. In these cases,
the number of free parameters are reduced by 1--3, and the low
$p_T$ range from 0 to 2.5--3.5 GeV/$c$ can be used.

The last two models use the relations between $T$ and $m_0$,
$\langle p_T \rangle$ and $\overline{m}$, as well as $\langle p
\rangle$ and $\overline{m}$. Due to the mass dependences of the
relations, these are not suitable to fit all particles
simultaneously in the low $p_T$ range. In principle, simultaneous
fits of all particles can be performed by using the first two
models. In the case of simultaneous fits, a larger $\chi^2$ can be
obtained due to the same set of parameters. Although we fit
different particle spectra by different sets of parameters, the
mean value of a given parameter can be obtained by weighting
different yields of the considered particles. Thus, the weighted
mean parameter can be regarded as a parameter suitable for the
simultaneous fit of all particles. Therefore, both the
simultaneous and non-simultaneous fits can be used in the analysis
of the particle spectra.

Based on the descriptions of the $p_T$ spectra, the first two
models can conveniently provide $T_0$ and $\beta_T$; however, the
values of parameters are possibly not the same according to
different models. To obtain the values of $T_0$, $\beta_T$, and
$\beta$ by models iii) and iv), we analyze the values of $T$
presented in Tables 3 and 4, and calculate $\langle p_T \rangle$,
$\langle p \rangle$, and $\overline{m}$ based on these values.
That is, we derived $\langle p_T \rangle$, $\langle p \rangle$,
and $\overline{m}$ by using a more complex fitting of Boltzmann
and Tsallis distributions in the $p_T$ range from 0 to $p_1$;
however, as the data was unavailable in certain regions the simple
counting of published spectra could not be used. Based on an
isotropic assumption in the rest frame of the emission source and
using a Monte Carlo method, we can perform the calculations
according to $p_T$ to obtain $\langle p \rangle$ and
$\overline{m}$ [15--17]. It can be seen that there are other
constraints in the statistical fits due to the excluding
contribution of the hard component and the selecting reference
frame of the emission source.

The relations between $T$ and $m_0$, $\langle p_T \rangle$ and
$\overline{m}$, as well as $\langle p \rangle$ and $\overline{m}$
are shown in Figs. 5, 6, and 7, respectively, where panels (a) and
(b) correspond to models iii) and iv) using Boltzmann and Tsallis
distributions, respectively. The symbols in Fig. 5 represent
values of $T$ listed in Tables 3 and 4 for different $m_0$ values.
The symbols in Figs. 6 and 7 represent values of $\langle p_T
\rangle$ and $\langle p \rangle$ for different $\overline{m}$
values, respectively, which are calculated from the parameters
listed in Tables 3 and 4 and by an isotropic assumption in the
rest frame of the emission source. The error bars in the three
figures represent overall errors. Although the method of least
squares was used to provide an appropriate connection, the lines
in the three figures connect the points for a better visibility in
each event sample. The intercept in Fig. 5 provides $T_0$, and
$\beta_T$ and $\beta$ can be obtained from the slopes in Figs. 6
and 7, respectively. The values of $T$, $T_0$, $\beta_T$, $\beta$,
and $\overline{m}$ can be considered independent of isospin.

\begin{figure*}[!htb]
\begin{center}
\includegraphics[width=16.0cm]{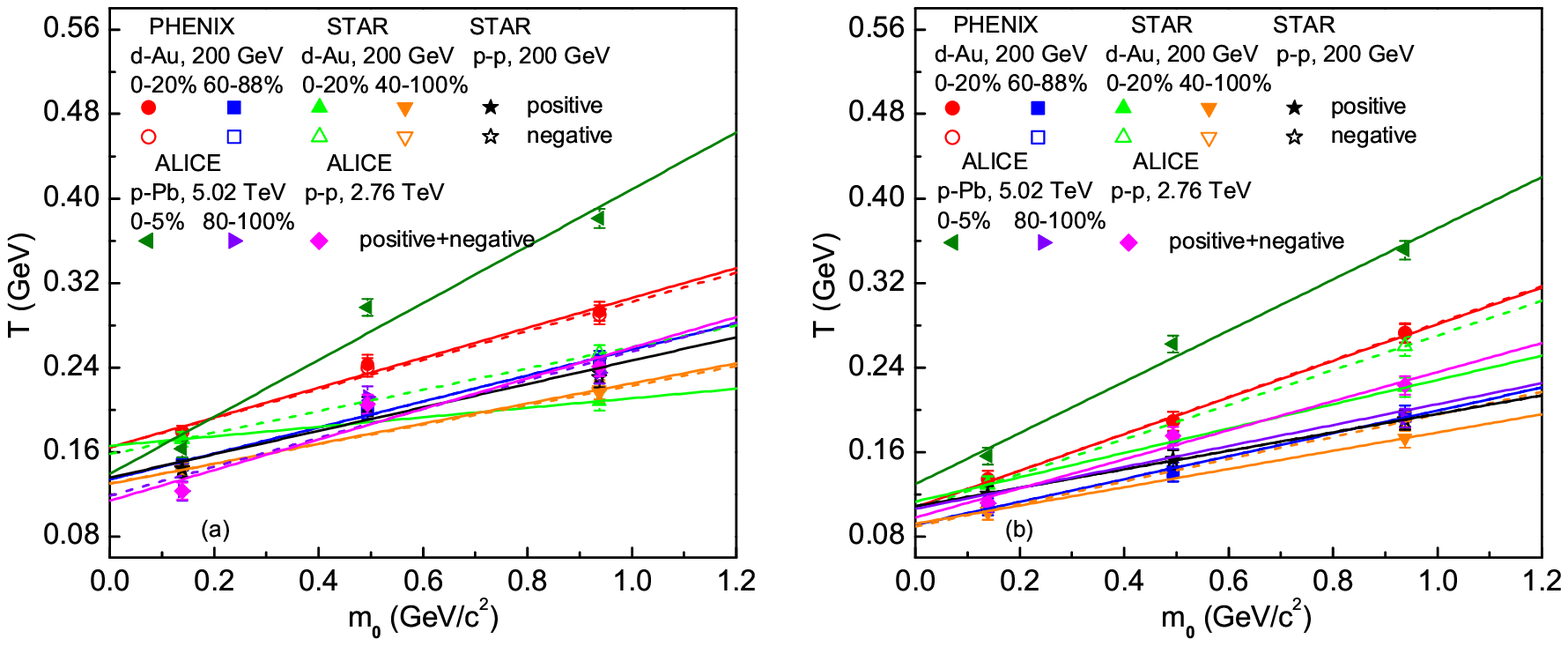}
\end{center} \vskip-.5cm
{\caption{\small (Color online) Relationship of $T$ and $m_0$,
where panels (a) and (b) correspond to models iii) and iv) using
Boltzmann and Tsallis distributions, respectively. The symbols
represent values of $T$ listed in Tables 3 and 4 for different
$m_0$ values. The lines connect the points for better
visibility.}}
\end{figure*}

\begin{figure*}[!htb] \vskip.5cm
\begin{center}
\includegraphics[width=16.cm]{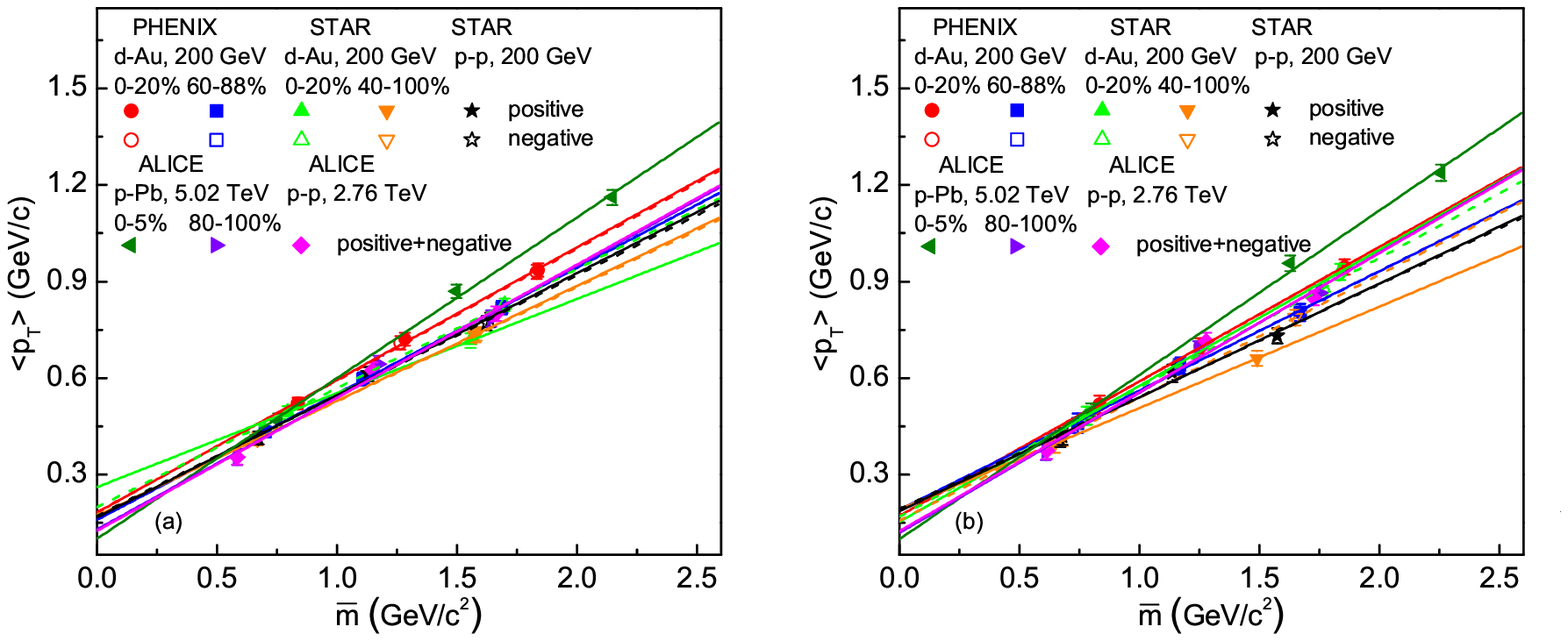}
\end{center} \vskip-.5cm
{\caption{\small (Color online) Relationship of $\langle p_T
\rangle$ and $\overline{m}$. The symbols represent values of
$\langle p_T \rangle$ for different $\overline{m}$ values,
calculated from the parameters listed in Tables 3 and 4 and by an
isotropic assumption in the rest frame of the emission source.}}
\end{figure*}

\begin{figure*}[!htb]
\begin{center}
\includegraphics[width=16.0cm]{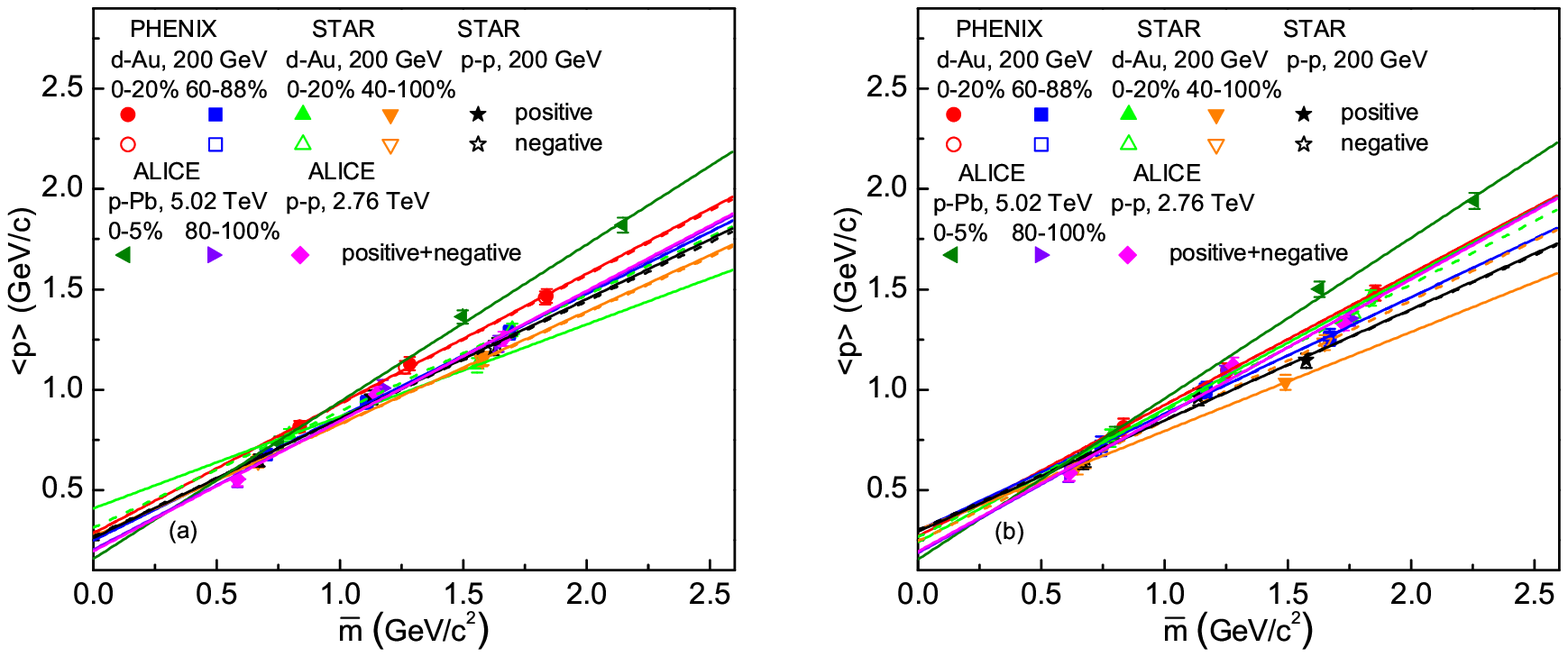}
\end{center} \vskip-.5cm
{\caption{\small (Color online) Relationship of $\langle p
\rangle$ and $\overline{m}$. The symbols represent values of
$\langle p \rangle$ for different $\overline{m}$ values,
calculated from the parameters listed in Tables 3 and 4 and by an
isotropic assumption in the rest frame of the emission source.}}
\end{figure*}

To compare values of key parameters obtained by different models
for different event samples, in the following we discuss the
qualitative dependences of $T_0$ and $\beta_T$ on the centrality.
From Tables 1 and 2, we can obtain $T_0$ and $\beta_T$ in the
first two models by weighting the yields of different particles.
From the intercept in Fig. 5, $T_0$ can be obtained in the last
two models, while from the slope in Fig. 6 (or 7), we can obtain
$\beta_T$ (or $\beta$) in the last two models. Generally, the four
models present similar results, and in certain cases these results
are in agreement with each other within errors. In central $d$-Au
and $p$-Pb collisions, $T_0$ is relatively greater than that in
peripheral collisions. Ranging from the RHIC to LHC energies,
$T_0$ shows a slight increase or the nearly invariant. Ranging
from the peripheral to central collisions and from the RHIC to LHC
energies, both $\beta_T$ show a slight increase or they are nearly
invariant. These conditions are in agreement with our recent work,
which studied Au-Au collisions at the RHIC and Pb-Pb collisions at
the LHC [14] by a slightly different superposition. In particular,
the absolute values of $T_0$ and $\beta_T$ do not show obvious
change in the range from $d$-Au ($p$-Pb) to Au-Au (Pb-Pb)
collisions, except for the systematical increase ($\le5$\%) due to
different superpositions. In $pp$ collisions, the dependences of
$T_0$ and $\beta_T$ on $\sqrt{s}$ are similar to those in
peripheral nuclear ($d$-Au, Au-Au, $p$-Pb, and Pb-Pb) collisions.

Apparently, in the above discussions $T_0$ and $\beta_T$ are
related to the physical properties of an expanding thermal system,
which, in the present work, is a high energy collision system with
a given impact parameter. If a mini-bias data sample is
considered, $T_0$ and $\beta_T$ are the averages over various
impact parameters. In particular, $T_0$ and $\beta_T$ in central
(peripheral) collisions are the averages over a given centrality
range. For $pp$ collisions without choosing a centrality, $T_0$
and $\beta_T$ are the averages over a given data sample and they
are related to the physical properties of the sample. In terms of
excitation degree, characterized by $T_0$, nuclear collisions such
as $d$-Au and Au-Au collisions at the RHIC and $p$-Pb and Pb-Pb
collisions at the LHC show similar excitation degree at the
kinetic freeze-out; however, the excitation degree in central
collisions is slightly higher than that in peripheral collisions.
The excitation degree depends on the heaviest nucleus, but
independent of the total nucleus, minimum nucleus, numbers of
participating nucleons, and binary collisions in nuclear
collisions at a given energy.

To confirm the above statement of the heaviest nucleus, instead of
using the total nucleus to determine $T_0$, in the following we
analyze copper-copper (Cu-Cu) collisions. Figure 8 shows the
spectra of $\pi^++\pi^-$, $K^++K^-$, and $p+\bar p$ produced in
0--20\% [Fig. 8(a)] and 40--94\% (60--92\%, 60--94\%, and
40--60\%) [Fig. 8(b)] Cu-Cu collisions at $\sqrt{s_{NN}}=200$ GeV.
The closed and open symbols represent the experimental data of the
PHENIX and STAR collaborations measured in $|\eta|<0.35$ and
$|y|<0.5$, respectively [29, 30], where the data of the 0--20\%
collisions are obtained by combining different centralities
(0--5\%, 5--10\%, and 10--20\%) to to match with those in Fig. 1,
and the data measured by different collaborations are connected by
scaling different amounts. The fit parameters are given in Tables
1--4, where the values of $N_0$ are obtained from the scaled
spectra, instead of the original spectra. It can be seen that the
four considered models approximately describe the $p_T$ spectra of
the identified particles produced in the central (0--20\%) and
peripheral (40--94\%) Cu-Cu collisions at $\sqrt{s_{NN}}=200$ GeV.

Figures 9(a) and 9(b) show the relationship of $T$ and $m_0$, as
well as $\langle p_T \rangle$ and $\overline{m}$ ($\langle p
\rangle$ and $\overline{m}$), according to the parameter values of
Cu-Cu collisions at $\sqrt{s_{NN}}=200$ GeV. It can be seen that
the mentioned relationship show nearly linear tendencies in most
cases. In particular, the intercept in Fig. 9(a) represents $T_0$,
and the slopes related to $\langle p_T \rangle$ and $\langle p
\rangle$ in Fig. 9(b) represent $\beta_T$ and $\beta$,
respectively.

For a qualitative comparison of the results obtained in different
types of collisions, Tables 1 and 2, as well as Figs. 5, 6 and 9
are examined by comparing with the values of $T_0$ and $\beta_T$.
It can be seen that the $T_0$ values in central Cu-Cu collisions
are slightly smaller than those in central $d$-Au (or $p$-Pb)
collisions because the size of Cu is smaller than that of Au (or
Pb). This is a direct and strong evidence for the statement that
the heaviest nucleus needs to be considered instead of the total
nucleus to determine $T_0$. In addition, the $T_0$ in peripheral
Cu-Cu collisions are nearly equal to those in peripheral $d$-Au
($p$-Pb) collisions and in $pp$ collisions. Apparently, the
dependence of $\beta_T$ on the size of heaviest nucleus is
undefined; however, $\beta_T$ in central collisions is comparable
with that in peripheral collisions.

\begin{figure*}[!htb] \vskip.5cm
\begin{center}
\includegraphics[width=16.0cm]{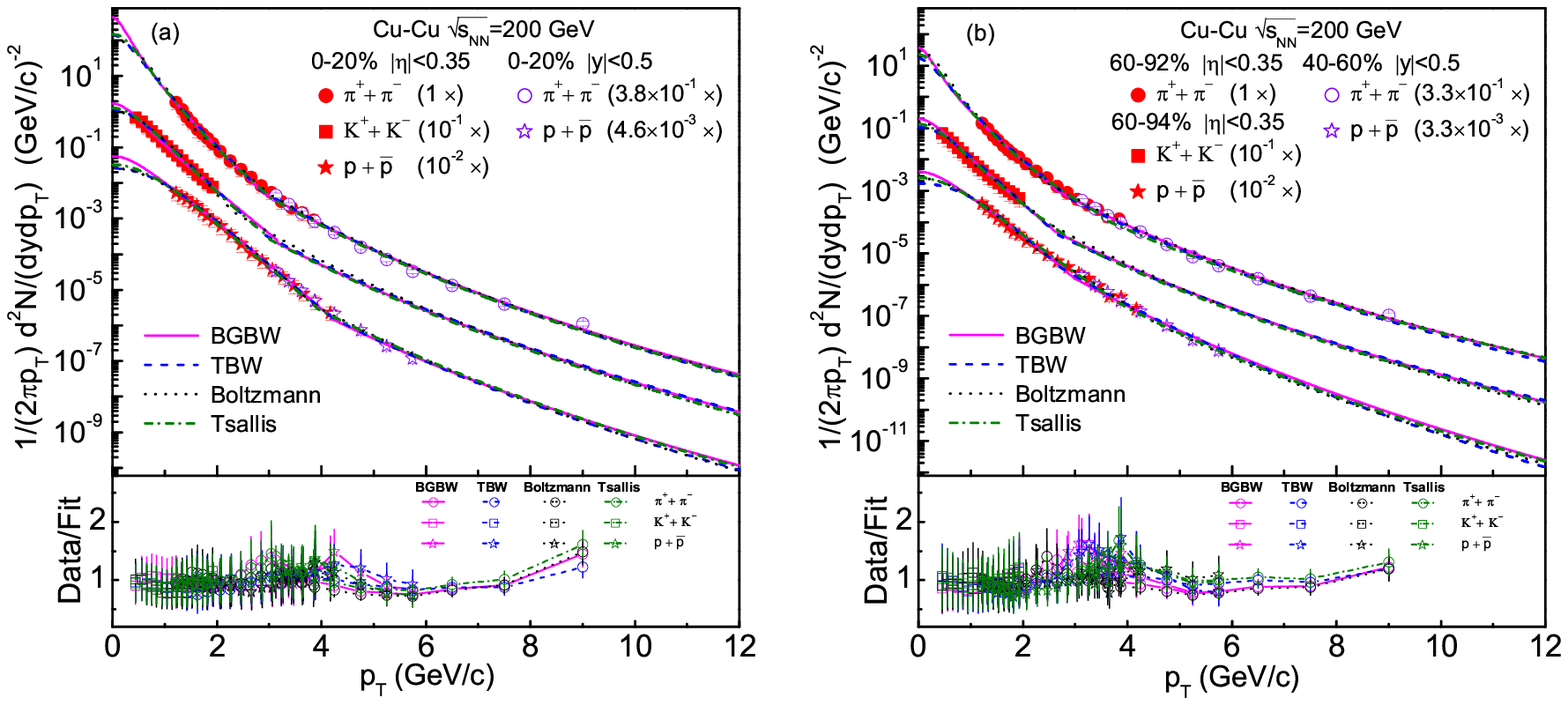}
\end{center} \vskip-.5cm
{\caption{\small (Color online) Spectra of $\pi^++\pi^-$,
$K^++K^-$, and $p+\bar p$ produced in 0--20\% [Fig. 8(a)] and
40--94\% [Fig. 8(b)] Cu-Cu collisions at $\sqrt{s_{NN}}=200$ GeV.
The closed and open symbols represent the experimental data of the
PHENIX and STAR collaborations measured in $|\eta|<0.35$ and
$|y|<0.5$, respectively [29, 30], where the data in 0--20\% were
obtained by combining different centralities (0--5\%, 5--10\%, and
10--20\%) to match with those in Fig. 1, and the data measured by
different collaborations are connected by scaling the different
amounts.}}
\end{figure*}

\begin{figure*}[!htb]
\begin{center}
\includegraphics[width=16.0cm]{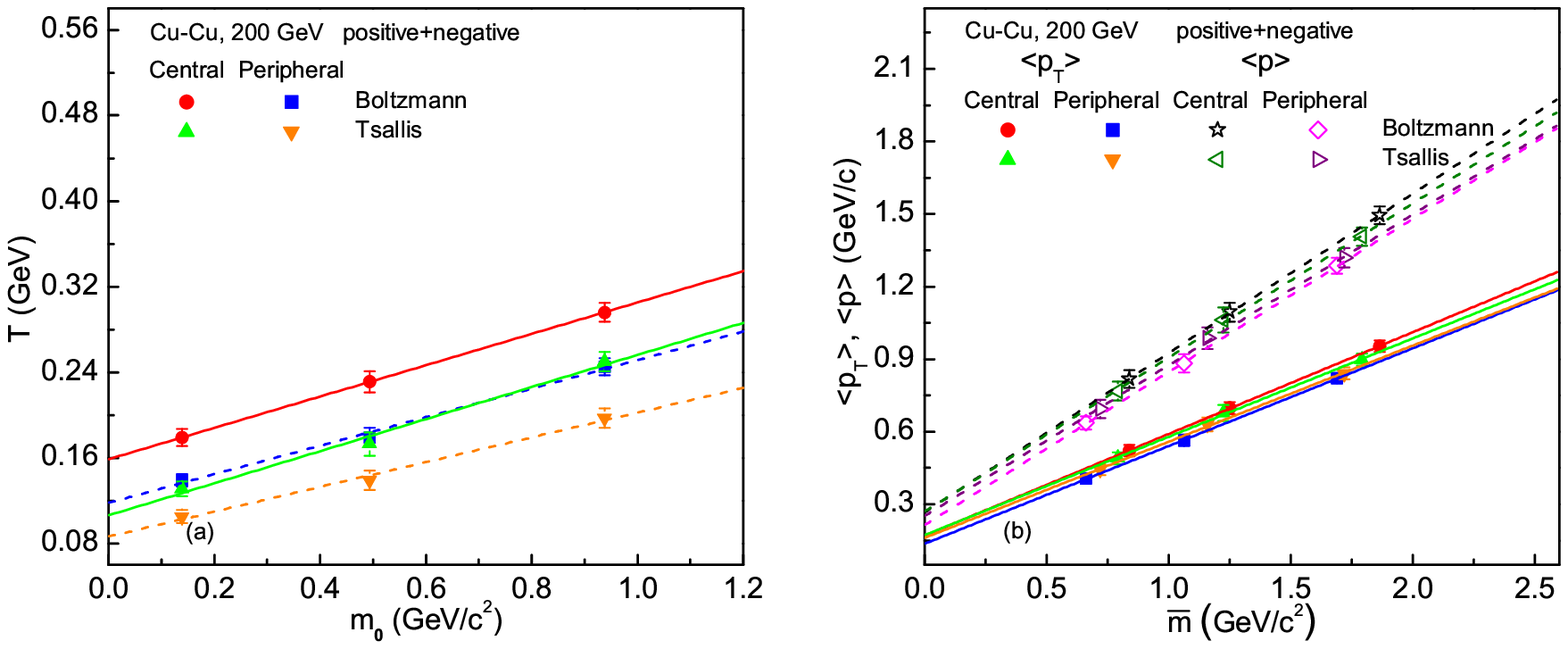}
\end{center} \vskip-.5cm
{\caption{\small (Color online) Relationship of (a) $T$ and $m_0$,
as well as (b) $\langle p_T \rangle$ and $\overline{m}$ ($\langle
p \rangle$ and $\overline{m}$), according to the parameter values
of Cu-Cu collisions at $\sqrt{s_{NN}}=200$ GeV.}}
\end{figure*}

The good agreement of the results obtained in the small system and
nucleus-nucleus collisions reveal certain universalities in the
hadroproduction process, as it is demonstrated in refs. [31-35].
The universality in the hadroproduction process appears in
different quantities observed [36] in different types of
collisions (including proton-proton, proton-nucleus, and
nucleus-nucleus collisions) and/or at different energies
(available in the range from SPS BES to LHC) [31--35]. These
quantities include, but not limited to mean multiplicity, rapidity
or pseudorapidity density, multiplicity or transverse momentum
distribution, and event patterns in different spaces under certain
conditions. The present work confirms that the universality in the
hadroproduction process possibly exists in thermal parameters at
kinetic freeze-out in different types of collisions ranging from
RHIC to LHC energies [14].

Although the blast-wave model and the related distributions have
no contributions from resonance decays and strong stopping
effects, a two-component form can be used to describe the spectra
in very-low and low $p_T$ ranges. In addition, in $d$-Au and
$p$-Pb collisions, the cold nuclear modification effects on the
$p_T$ spectra are not considered by us; however, a few of them
widening the $p_T$ spectra of the identified particles due to the
multiple cascade collisions in the cold spectator region. If the
contribution of the effects of cold nuclear modification on the
$p_T$ spectra is excluded, smaller $T_0$ and $\beta_T$ can be
obtained. The comparison with $pp$ collisions reveals that the
contribution of cold nuclear modification effects on $T_0$ and
$\beta_T$ is not obvious because peripheral nuclear collisions and
$pp$ collisions have similar values. Furthermore, the contribution
of cold nuclear modification effects on $T_0$ and $\beta_T$ in
Au-Au (Pb-Pb) collisions at the RHIC (LHC) is not obvious as well
[14].

The Tsallis function is connected to the thermal model via its
fits to the two- or three-component Boltzmann distribution [37].
Index $q$ represents the degree of non-equilibrium among two or
three states described by Boltzmann distributions, and the Tsallis
temperature describes the fluctuations of Boltzmann temperatures.
These explanations on the level of drawing curves of $p_T$ reveal
that the interacting systems at the RHIC and LHC stays in a
transitional region from the extensive system to the non-extensive
system. There is no obvious boundary to distinguish the extensive
system and the non-extensive system for a given interacting system
in the considered energy range. Nevertheless, at the RHIC and its
beam energy scan energies or similar energies, the generic
axiomatic non-extensive statistics is used to obtain the chemical
freeze-out temperature and the baryon chemical potential [38--40].
This indicates that the Boltzmann-Gibbs and Tsallis statistics are
not always necessary or applicable, which suggest that the
interacting systems at the considered energies are complex, and
more studies are needed in the future.

In central collisions at RHIC and LHC energies, the kinetic
freeze-out temperature obtained from the four models is
$T_0\approx120$ MeV. It is lower than the chemical freeze-out
temperature $T_{ch}\approx160$ MeV [1--4]. This confirms that the
kinetic freeze-out occurs later than the chemical freeze-out at
the considered energies. As an approximate treatment, we consider
an ideal fluid, in which the time evolution of the temperature
follows $T_f=T_i(\tau_i/\tau_f)^{1/3}$, where $T_i$ and $\tau_i$
are the initial temperature and proper time, respectively [41,
42], and $T_f$ and $\tau_f$ denote the final temperature and time,
respectively. When considering $T_i=300$ MeV and $\tau_i=1$ fm
[42], the chemical freeze-out occurs at $\tau_{ch}\approx 6.6$ fm
and the kinetic freeze-out occurs at $\tau_0\approx 15.6$ fm. When
considering peripheral collisions, the kinetic freeze-out occurs
at $T_0\approx105$ MeV and $\tau_0\approx23.3$ fm. For instance,
if a non-ideal fluid is considered, the viscosity to entropy
density ratio $\eta/s$ is considered as 0.2, the time delay for
the two freeze-outs is small, compared with the ideal fluid.

Let us summarize the main contributions of the present work as
follows. Before reconsidering the first two models, applying a
nearly zero $\beta_T$ in them, the four models do not exhibit
similar results. After reconsidering the first two models,
applying a non-zero $\beta_T$ in them, the four models exhibit
similar results. By comparing the central nuclear collisions, the
proton-proton collisions are found to be closer to the peripheral
nuclear collisions, especially in terms of $T_0$ and $\beta_T$.
The $T_0$ ($\beta_T$) value in the central collisions is
comparable with that in the peripheral collisions, and $T_0$
($\beta_T$) value in collisions at the LHC is comparable with that
at the RHIC. At any rate, $T_0$ ($\beta_T$) value in the central
collisions is not smaller than that in the peripheral collisions,
and $T_0$ ($\beta_T$) at the LHC is not smaller than that at the
RHIC.

Before the final conclusions, it should be emphasized that the
comparisons of different models and the obtained $T_0$ and
$\beta_T$ values in small collision system presented in this study
are significant and useful owing to the collective expansion in a
small system [43]. This also indicates that a large $\beta_T$
($\sim0.4c$) is applied in peripheral nuclear collisions and $pp$
collisions. As we know, certain models [6--8, 44--52] are used to
obtain $T_0$ and $\beta_T$, and it is difficult to obtain the
similar results compared to others [53--59] from these models with
the increase of quantities. Although the present work provides
similar results to [53--59] by the four models, the first and
third models are preferred as they use a Boltzmann distribution,
which is closer to the well-known ideal gas model. In addition,
the hard component has no contribution to $T_0$ and $\beta_T$ due
to its non-thermal production. Instead, the very-soft and soft
components which contribute fitly in the very-low and low $p_T$
regions, are used to obtain $T_0$ and $\beta_T$. Thus, the third
and fourth fits are suitable, because they can be applied for
massive particles and in very-low and low $p_T$ ranges.

In addition, complex physics processes, high energy collisions
contain abundant information. This information includes, but is
not limited to, electromagnetic field effects [60], strong
magnetic field effects [61], and particular effects of strangeness
[62]. The determination of $T_0$ and $\beta_T$ can be affected by
these effects; hence, the search for the QCD critical point [63].
As a study at the exploratory stage of development, the present
work still has needs to be improved with the highest possible
accuracy. Further studies needs to be focused on the accurate
determination of $T_0$ and $\beta_T$. In addition, the accurate
determination of other types of temperatures, such as the
effective temperature, chemical freeze-out temperature, and
initial temperature, and comparisons of their dependences on the
centrality and collision energy is also in the focus of our
research.
\\

{\section{Conclusion}}

As a conclusion, the transverse momentum distributions of $\pi^+$,
$\pi^-$, $K^+$, $K^-$, $p$, and $\bar p$ produced in $pp$ and
$d$-Au collisions at the RHIC, as well as in $pp$ and $p$-Pb
collisions at the LHC, have been analyzed by four models. The
first two models utilize the blast-wave model with Boltzmann-Gibbs
statistics and with Tsallis statistics, respectively. The last two
models employ certain linear correspondences, in which the
Boltzmann and Tsallis distributions are used to obtain the
effective temperatures. These models and distributions describe
only the contribution of the soft excitation process. For the hard
scattering process, the inverse power law is uniformly used.

The experimental data measured by the PHENIX, STAR, and ALICE
collaborations are fitted by the model results. We used a non-zero
$\beta_T$ in the first two methods. The four models present
similar results. Both $T_0$ and $\beta_T$ in central collisions
are comparable with those in peripheral collisions. With the
increase of collision energy ranging from that of the RHIC to that
of the LHC, the considered quantities typically do not decrease.
Comparing with the central nuclear collisions, the $pp$ collisions
are closer to the peripheral nuclear collisions. In nuclear
collisions, the excitation degree at the kinetic freeze-out is
mainly determined by the heaviest nucleus and collision energy.
\\

{\bf Conflicts of Interest}

The authors declare that there is no conflict of interests
regarding the publication of this paper.
\\

{\bf Acknowledgments}

Communications with Muhammad Waqas are highly appreciated. This
work was supported by the National Natural Science Foundation of
China under Grant Nos. 11575103 and 11747319, the Shanxi
Provincial Natural Science Foundation under Grant No.
201701D121005, the Fund for Shanxi ``1331 Project" Key Subjects
Construction, and the US DOE under contract
DE-FG02-87ER40331.A008.
\\

\end{multicols}
\end{document}